\documentclass[conference,letterpaper]{IEEEtran}

\usepackage[utf8]{inputenc} 
\usepackage[T1]{fontenc}
\usepackage{url}
\usepackage{ifthen}
\usepackage{cite}
                             
\usepackage[cmex10]{amsmath}
\usepackage{amssymb}
\usepackage{algorithm}
\usepackage{algorithmic}
\usepackage{graphicx}
\usepackage{textcomp}
\usepackage{xcolor}
\usepackage{ulem}
\usepackage{cancel}
\usepackage{physics}
\DeclareMathOperator*{\argmax}{arg\,max}
\DeclareMathOperator*{\argmin}{arg\,min}
\newtheorem{theorem}{Theorem}
\usepackage{enumitem}
\newcommand{\remove}[1]{}

\IEEEoverridecommandlockouts 

\begin{document}
\title{Greedy $k$-Center from Noisy Distance Samples} 

\author{%
  \IEEEauthorblockN{Neharika Jali, Nikhil Karamchandani, and Sharayu Moharir}
  \IEEEauthorblockA{Department of Electrical Engineering\\
                    Indian Institute of Technology, Bombay\\
                    Email: neharikajali@iitb.ac.in, \{nikhilk, sharayum\}@ee.iitb.ac.in}
}

\maketitle

\begin{abstract}
 We study a variant of the canonical \textit{$k$-center problem} over a set of vertices in a metric space, where the underlying distances are apriori unknown. Instead, we can query an oracle which provides noisy/incomplete estimates of the distance between any pair of vertices. We consider two oracle models:  \textit{Dimension Sampling} where each query to the oracle returns the distance between a pair of points in one dimension; and \textit{Noisy Distance Sampling} where the oracle returns the true distance corrupted by noise. We propose active algorithms, based on ideas such as UCB, Thompson Sampling and Track-and-Stop developed in the closely related Multi-Armed Bandit problem, which adaptively decide which queries to send to the oracle and are able to solve the $k$-center problem within an approximation ratio of two with high probability. We analytically characterize instance-dependent query complexity of our algorithms and also demonstrate significant improvements over naive implementations via numerical evaluations on two real-world datasets (Tiny ImageNet and UT Zappos50K).
\end{abstract}

\section{Introduction}
Many problems in machine learning and signal processing are based on computing and processing distances over large datasets, for example clustering and nearest-neighbor decoding. In several applications, the algorithm might only have access to noisy estimates of the underlying pairwise distances. For example, the distance between two locations in a city being estimated using travel times, or the similarity between items on an e-commerce portal measured using ratings/feedback given by users. In this work, we consider the canonical \textit{$k$-center problem} where given some $n$ points in a metric space, the goal is to find the subset of $k$ centers such that the maximum distance of any point to its nearest center is minimized. While this problem is NP-hard, a simple greedy algorithm, which operates in phases and adds the farthest point from the current set of centers in each phase,  achieves an approximation factor of two if all the pairwise distances are available \cite{KT}. Instead, we assume that the algorithm does not know the distances apriori, but has access to an oracle which it can query to get partial/noisy information about the distance between any pair of points. Our goal is to design sequential learning algorithms which can approximately solve the $k$-center problem, while aiming to minimize the total number of queries required by adaptively selecting queries  based on earlier responses.

Specifically, we consider the  two query models: \textit{Dimension Sampling} where the points are assumed to lie in a high-dimensional space and each query to the oracle returns the distance between a pair of points along a particular dimension. This sampling model is motivated by recent work in \cite{Bagaria} and \cite{LeJeune} which use random dimension sampling to significantly reduce the computational complexity of problems such as nearest neighbor graph construction and $k$-means clustering when applied to huge datasets with applications such as single-cell RNA sequencing ($10\times$ Genomics 68k PBMCs scRNA-seq dataset); and \textit{Noisy Distance Sampling} where upon being queried with a pair of points, the oracle returns the true distance corrupted by some additive noise. Such a model has natural applications in several settings, for example in network tomography where delay measurements are used to provide an estimate of the distance between servers \cite{tsang2004network}; or in preference learning where user preferences can be modeled as distances with the smallest distance representing the most preferred item \cite{kruskal1964nonmetric, shepard1962analysis, jain2017learning} and user activity such as ratings for similar items providing noisy estimates of the user-item distances. The nearest neighbor problem under a similar oracle model was recently studied in \cite{MasonAndTripathy, mason2021nearest}. 

\section{Our contributions}

We make the following contributions towards understanding the query complexity of solving the $k$-center problem. 

\textit{(a)} Under each of the above query models, we design sequential learning algorithms which mimic the greedy algorithm with access to the true pairwise distances and return with large probability the same set of $k$ centers, and thus achieve the same approximation factor of two. Furthermore, we provide \textit{instance-dependent} upper bounds on the query complexity of these algorithms which explicitly depends on the various pairwise distances in the dataset. Our bounds indicate that there might be some instances where solving the $k$-center problem is `easy' in that it requires few queries, for example if the {data points are reasonably spaced} and are naturally divided into $k$ far-apart clusters; whereas there can also be `hard' instances where many more queries are needed to distinguish between several close candidates for the centers. Our proposed algorithms are based on ideas such as upper and lower confidence bounds (UCB, LCB), Thompson Sampling and Track-and-Stop \cite{Lattimore, kaufmann16a} from online learning, and use these to effectively guide adaptive exploration across rounds. 

\textit{(b)} For a restricted class of datasets and algorithms, we derive fundamental lower bounds on the query complexity of solving the $k$-center problem and find that their expressions closely resemble the query complexity upper bounds of our proposed schemes. This indicates that the algorithms correctly assign oracle queries among the various node pairs and make good use of the information contained in the responses towards identifying the $k$ centers.  

\textit{(c)} We demonstrate the superior performance of our proposed schemes over other baselines via extensive simulations on two real datasets ( Tiny ImageNet \cite{tinyimagenet} and UT Zappos50K \cite{zappos1}, \cite{zappos2}).

\color{black}

\section{Related Work}
The $k$-center problem is a classical combinatorial optimization problem with numerous applications ranging from facility location \cite{Thorup} to clustering \cite{AgarwalClustering, Badoiu} and data/graph analytics \cite{Abbar, Ceccarello}. The problem is known to be NP-hard in general, but several approximation algorithms exist \cite{Gonzalez, FederandGreene, Dorit}. In particular, \cite{Gonzalez} designed a greedy algorithm which achieves an approximation factor of two and showed that this factor is the best possible if $P\neq NP$. While this line of work assumes that the points (and hence the distances) are available exactly, our focus in this work is on the design of sequential learning algorithms which adaptively query an oracle whose responses are partial/noisy versions of the pairwise distances. Though there has been work on online and streaming algorithms for solving the $k$-center problem \cite{CeccarelloVLDB, Charikar}, the focus there is on optimizing resource consumption such as memory and computational effort. 

Modern data science applications often involve computations over datasets that  contain not only a large number of entries but also a large number of dimensions or features. Such problems can often have huge computational complexity and there are several recent works which have proposed the use of randomized adaptive algorithms for significantly speeding up a variety of computational problems. These include Monte Carlo tree search for Markov decision processes in \cite{KandS}, hyperparameter tuning for deep neural networks in  \cite{Li} and \cite{JandT}, and discrete optimization problems in \cite{Bagaria}.  
Works which are closer to the theme of this paper include \cite{Bagaria} and \cite{LeJeune} which studied the $k$-nearest neighbor problem in high dimensions and designed algorithms which adaptively estimate distances between selected pairs of nodes by sampling dimensions randomly. These fall under the purview of the Dimension Sampling model studied in this paper. \cite{MasonAndTripathy, mason2021nearest} studied the sample complexity of sequential algorithms for learning nearest neighbor graphs under a query model very similar to the Noisy Distance sampling model considered in our work here. \cite{Bagaria2}, \cite{BaharavandTse} use adaptive sampling to reduce the complexity of identifying the medoid among a collection of points, while \cite{Tiwari} very recently extended these ideas to speed up the very popular Partitioning Around Medoids (PAM) algorithm used for $k$-medoid clustering. Finally, \cite{addanki2021design} recently studied the problem of $k$-center and hierarchical clustering using a different oracle model that compares the relative distance between the queried points and provides noisy answers to questions of the form `Is point $u$ closer to $v$ or $w$ closer to $x$?'.

Adaptive decision making is a central theme of the well-known multi-armed bandit problem, see for example \cite{Lattimore}. Ideas such as confidence intervals and Bayesian sampling for designing adaptive schemes and change of measure arguments for deriving lower bounds on the sample complexity are common in the bandit literature, and find application in our work as well as most of the works mentioned above. In particular, mapping each pair of points to an arm with an expected reward equal to their pairwise distance, each stage of the greedy algorithm for the $k$-center problem can be thought of as a best-arm identification problem with a max-min reward objective instead of a max reward objective, as is standard in the pure exploration framework in bandit literature \cite{Jamieson}. { \cite{garivier2016maximin} is one of the few works, to the best of our knowledge, that studies the max-min reward objective.} This fundamentally changes the nature of efficient schemes and query complexity bounds, and requires careful adaptation even when applying confidence intervals and change of measure arguments. For example, in Section~\ref{Sec:DS-TS}, we propose a Thompson Sampling based scheme and argue why it should perform better than the UCB sampling based alternative (demonstrated via simulations later). On the other hand, Thompson sampling based algorithms are known to perform poorly for best arm identification under the standard max reward objective. 

Finally, while the works discussed above along with our work here are based on adaptively estimating distances, there exist several other approaches for overcoming the complexity in high dimensions such as using random projections \cite{Ailon}, locality sensitive hashing \cite{AndoniandIndyk}, and ordinal embedding \cite{Jain}. These approaches often involve pre-processing steps whose complexity becomes non-negligible if either the number of tasks executed is small and hence the cost cannot be amortized, or if the underlying dataset changes rapidly over time. 

The rest of the paper is organized as follows. Section~\ref{section:problem} describes the problem setting including the two oracle models. Section~\ref{Sec:LowerBound} establishes a lower bound on the query complexity for a restricted class of algorithms. Next, Sections~\ref{Sec:DS} and \ref{Sec:NS} describe various algorithms for the two oracle models and also provide query complexity analysis. In Section~\ref{Sec:Experiments}, we provide extensive experimental validation of our policies on two real-world datasets. Finally, we present some conclusions in Section~\ref{Sec:Conclusions}.

\section{Problem Setup}
\label{section:problem}
 Consider a set of $n$ points, $\mathcal{V} =\{x_1, x_2, \dots, x_n\} \in \!\mathbb{R}^m$. Let $d_{u,v}$ represent the normalized squared distance\footnote{While we focus on $\ell_2$ distance here, our results and ideas generalize to other distance measures { that are separable}.} between points $u$ and $v$ as
\begin{align*}
    d_{u,v} = \frac{||x_u - x_v||_2^2}{m} = \frac{\sum\limits_{i=1}^m |[x_u]_j - [x_v]_j|^2}{m} \overset{\Delta}{=} \frac{\sum\limits_{i=1}^m [d_{u,v}]_j}{m} 
\end{align*}
where $[x]_j$ represents the $j$-th coordinate {of} point $x$. { For any subset $\mathcal{U} \subset \mathcal{V}$, we define  the bottleneck distance as}
\begin{equation} 
\Delta(\mathcal{U}) = \max_{v \in \mathcal{V}\backslash\mathcal{U}} \min_{u \in \mathcal{U}} d_{v, u}.
\end{equation}
Define $\mathcal{Y}$ to be the set of all subsets $\mathcal{S}$ with exactly $k$ points. The objective of the $k$-center problem is to find the subset $\mathcal{S}$ with the smallest value of $\Delta(\mathcal{S})$, i.e.,
\begin{equation} 
\mathcal{S}^{\star} = \argmin_{\mathcal{S} \in \mathcal{Y}} \Delta(\mathcal{S}).   
\end{equation}
{ For convenience we assume that all points are normalized such that $||x||_{\infty} \leq 1/2$, where $||x||_{\infty}$ denotes the largest absolute value of the vector $x$.}

 The $k$-center problem is known to be NP-hard \cite{KT} and can be solved approximately by a greedy algorithm  \cite{Gonzalez}. The algorithm begins by choosing an arbitrary point as the first center, say $s_1$. The following step is then repeated until we find $k$ centers. If the current set of centers is $\mathcal{S}_i = \{s_1, s_2, \dots, s_i\}$, the next center $s_{i+1}$ is chosen as the point $v \in \mathcal{V}\backslash\mathcal{S}_i$ that achieves\footnote{Here, we assume that in any step, a unique point achieves this bottleneck distance. While this imposes a slight limitation on the algorithms presented below, this almost always holds true for real life datasets. For cases which do have more than one such optimal point, more robust versions of the algorithm can be designed along the lines of the $\epsilon$-best arm identification problem \cite{kgepsilonbest} in multi-armed bandits.} the bottleneck distance, $\Delta(\mathcal{S}_i)$, i.e.,   
 \begin{equation}
    s_{i+1} = \max_{v \in \mathcal{V}\backslash\mathcal{S}_i} \min_{u \in \mathcal{S}_i} d_{v, u} . 
    \label{eqn:greedystage}
 \end{equation}
 If $G$ is the set of centers found by the greedy algorithm, $O$ is the set of optimal centers, and the distance measure follows the triangle inequality, it is guaranteed that $\Delta(G) \leq 2\Delta(O)$. The greedy algorithm as described above requires knowledge of the exact distances between each pair of points. We consider the $k$-center problem when the algorithm { does not} have direct access to such information and can only query a noisy oracle. We consider the following two frameworks for queries which are used to estimate the pairwise distances.
\begin{itemize}[leftmargin=*]
    \item \textbf{Dimension Sampling {(DS)}}: When the oracle is queried with the triplet $(u,v,j)$ where $u,v \in \{1,2, \dots, n\}$ and $j \in \{1,2, \dots, m\}$, its response $\mathcal{O}(u,v,j)$ is the square of the distance between the pair of points $(x_u, x_v)$ in dimension $j$, {denoted by $[d_{u,v}]_j$} i.e.,
    \begin{equation}\label{eqn:DSModel}
        \mathcal{O}(u,v,j) = [d_{u,v}]^2_j = |[x_u]_j - [x_v]_j|^2.
    \end{equation}
    \item \textbf{Noisy Distance Sampling {(NS)}}: For each query $(u,v)$, the oracle returns an independent noisy observation of the true distance $d_{u,v}$ between $x_u$ and $x_v$. In particular, 
    \begin{equation}\label{eqn:NSModel}
        \mathcal{O}(u,v) = d_{u,v} + \eta.
    \end{equation}
    where $\eta$ is a zero mean, variance $\sigma^2_{noise}$ Gaussian random variable. We assume the noise\footnote{The noise distribution can be generalized easily to sub-Gaussian distributions owing to the existence of equivalent concentration inequalities.}  to be i.i.d. across queries. 
\end{itemize}

 The goal of this work is to design sequential algorithms for the $k$-Center problem that mimic the stages of the greedy algorithm, while using fewer queries by adaptively estimating distances. We first describe the framework we adopt to solve this problem. Conceptually, we map the problem to the multi-armed bandit (MAB) framework where each pair of points $(x_v, x_s)$ corresponds to an arm $a_{v,s}$. When the arm $a_{v,s}$ is pulled, the reward $r_{v,s}$ is obtained. In case of the Dimensional Sampling model, we propose algorithms which sample a random dimension for any pair of nodes they query, and thus the random reward corresponds to the squared distance $[d_{v,s}]^2_j$ between the points $x_v$ and $x_s$ in a uniformly random chosen dimension $j$, given by $r_{v,s} = \mathcal{O}(v,s,j) = |[x_v]_j - [x_s]_j|^2$.
In case of the Noisy Distance Sampling model, the reward corresponds to noisy observation of the true distance which is  $r_{v,s} = \mathcal{O}(v,s) = d_{v,s} + \eta$. It is easy to verify under both the models that the expected reward of an arm $a_{v,s}$ is equal to the true normalized squared distance and distance value respectively. The objective of the greedy algorithm in any stage $i$ then maps to finding the arm corresponding to the max-min expected reward
\begin{equation}
    \max_{v \in \mathcal{V}\backslash\mathcal{S}_i} \min_{s \in \mathcal{S}_i} \mathbb{E}[ r_{v,s} ] =  \max_{v \in \mathcal{V}\backslash\mathcal{S}_i} \min_{s \in \mathcal{S}_i} d_{v,s} .
    \label{Eqn:MABReward}
\end{equation}
Notice that unlike the maximization objective in the standard best-arm identification framework \cite{Jamieson}, we have a max-min objective here. {The max-min reward objective, thought not common, has been previously studied in \cite{garivier2016maximin}.}
\color{black}

\section{Lower Bound}\label{Sec:LowerBound}
In this section, we present a lower bound on the query complexity of the $k$-center problem for a restricted class of datasets and algorithms under the Dimension Sampling model. {\color{black} We will restrict attention to the class of algorithms which proceed in stages $1,2,\ldots, k$, identifying a new center $o_i$ in each stage $i$. For such an algorithm, we will say that it `mimics' the output of the greedy algorithm for any dataset if in each stage $i \in \{1,2,\ldots,k\}$, the output $o_i$ matches the center $s_i$ identified by the greedy algorithm as given by \eqref{eqn:greedystage}. Note that both the identities and the sequence in which the centers are identified matters here.} Deriving a more general lower bound is open even for simpler problems such as the nearest neighbor computation (which corresponds to $k=2$ centers) for which similar restrictions were used in \cite{LeJeune}. For any stage $p$ of the scheme, we have an MAB setting that can be thought of as each node corresponding to a box and containing multiple arms, one for the distance to each of the current centers. The score associated with a box is the minimum expected reward of its arms and the goal is to find the box with the highest score. This is a generalization of the standard best arm identification problem, which we think is interesting in its own right. 
\begin{theorem} \label{lowerboundtheorem}
For $\delta \in (0,1)$ {\color{black} and for any dataset $\mathcal{V}$}, consider an algorithm $\mathcal{A}$ that can obtain information by only observing pairwise distances across randomly chosen coordinates and {\color{black} proceeds in stages $1,2,\ldots, k$, identifying a new center $o_i$ in each stage $i$. Furthermore, assume that with probability at least $(1- \delta)$, $\mathcal{A}$ mimics the output of the greedy algorithm so that for each stage $i \in \{1,2,\ldots,k\}$, the output $o_i$ matches the center $s_i$ identified by the greedy algorithm as given by \eqref{eqn:greedystage}.} Then, when algoithm $\mathcal{A}$ is applied to a dataset where $[x]_j \in \{\frac{1}{2}, -\frac{1}{2}\}$ for every point $x$, the expected number of queries required is at least 
\begin{align}
    \mathbb{E}[T] \geq & \frac{1}{2} \sum\limits_{v' \in \mathcal{V} \backslash \mathcal{S}_k} \max\limits_{p=\{1,\dots,k-1\}} \frac{\log(1/2.4\delta)}{\max\limits_{i=\{1,\dots,p\}}kl_b(d_{v',s_i}, d_{s_{p+1}, s_i})}  \nonumber\\
    & + \frac{1}{2} \sum\limits_{p=1}^{k-1} \max\limits_{v'' \in \mathcal{V} \backslash \mathcal{S}_{p+1}} \frac{\log(1/2.4\delta)}{\max\limits_{i=\{1,\dots,p\}}kl_b(d_{s_{p+1}, s_i}, d_{v'',s_i})} ,\nonumber
\end{align}
where {$kl_b(x,y)$} represents the KL divergence between two Bernoulli distributions with means $x$ and $y$. 
\end{theorem}

Recall that $s_i$ and $\mathcal{S}_p$ denote the $i$-th center and the set of first $p$ centers respectively, as identified by the greedy algorithm for the underlying instance. In the above result, we assume all coordinates for any point in the dataset to be in $\{-1/2 ,1/2\}$. This restriction was used in \cite{LeJeune} as well and is done to match the support set for all observations, thus ensuring that the $KL$-divergence terms are well defined. In the above expression of the lower bound, each term in the first summation corresponds to a sub-optimal point $v'$ at a particular stage $p$, and the $KL$-divergence term compares for each existing center $s_i$, its distance to the sub-optimal point $v'$ with that to the  optimal point in this stage denoted by $s_{p+1}$. The inverse dependence on the $KL$-divergence terms agrees with the intuition that the hardness of reliably identifying the next center is higher when the distances corresponding to a sub-optimal point and the optimal point are closer. The second summation in the expression corresponds to the set of centers and follows a similar principle. The proof for the above theorem is presented in Appendix \ref{lowerboundproof}. A  lower bound for the Noisy Distance Model can be derived along very similar  lines. In the following sections, we will design sequential algorithms for finding the $k$-center in both the query models and compare their query complexity to the lower bound derived above. 
\remove{
Intuitively, each term in the set of terms over which the summation exists is associated with the best arm in the box corresponding a sub-optimal point. The last term corresponds to the best arm in the box corresponding to the optimal point, which is identified as the next center in a particular round. The presence of terms corresponding to only one arm from each box in this lower bound gives us a clue about the nature of efficient policies. We build upon these hints to design algorithms in the following sections.}
\section{Dimension Sampling Model}\label{Sec:DS}
 The greedy algorithm for the $k$-center problem discussed in Section~\ref{section:problem} proceeds in stages, adding one new center in each stage. A naive implementation\footnote{The naive implementation of the greedy approach is called the naive greedy algorithm hereafter.} of such an algorithm under the Dimension Sampling {(DS)} oracle model will compute all the relevant distances exactly in each stage by querying all the dimensions one by one. The query complexity of such a 
brute-force algorithm will be $O(nmk)$ which is prohibitively high for high-dimensional data. We instead design an algorithm which adaptively learns distance estimates and can have significantly lower query complexity. 

\subsection{DS-UCB}
\subsubsection{DS-UCB: Algorithm}
Drawing inspiration from the popular Upper Confidence Bound (UCB) algorithm \cite{LaiandRobbins}, we propose the DS-UCB (Algorithm~\ref{algo:dsucb}) scheme which can provide significant savings over the naive greedy algorithm in terms of the number of dimensions sampled between a pair of points. 

During the course of the scheme, we may pull an arm $a_{v,s}$ multiple times and say after $t_{v,s}$ pulls, estimate the corresponding distance $d_{v,s}$ using the unbiased estimator $\hat{d}_{v,s}(t_{v,s})$, given by 
\begin{equation}
    \hat{d}_{v,s}(t_{v,s}) = \sqrt{\frac{1}{t_{v,s}} \sum\limits_{j \in j_1,...,j_{t_{v,s}}} |[x_v]_j - [x_s]_j|^2}
\end{equation}
where $j_i$ denotes the dimension supplied to the oracle as input in the $i$-th pull. Furthermore, using a non-asymptotic version of law of iterated logarithm \cite{jamieson2014lil, KaufmannLIL}, we know that with probability at least $(1-\delta')$ the true distance $d_{v,s}$ lies within a confidence interval of size $\alpha(t_{v,s}, \delta')$ where
$$
 \alpha_{v,s}(t_{v,s}, \delta')  \propto \sqrt{\frac{\log(\log (t_{v,s})/ \delta')}{t_{v,s}}}.
$$

Upon the next pull of the arm $a_{v,s}$, we update\footnote{We hereafter abbreviate $\hat{d}_{v,s}(t_{v,s})$ as $\hat{d}_{v,s}$ and $\alpha_{v,s}(t_{v,s},\delta')$ as $\alpha_{v,s}$.} the distance estimate $\hat{d}_{v,s}$, the upper confidence bound (UCB) { $U(d_{v,s}) = \hat{d}_{v,s} + \alpha_{v,s}$} and the lower confidence bound (LCB) { $L(d_{v,s}) = \hat{d}_{v,s} - \alpha_{v,s}$}.  
For any point $v$, let $s_v^{\star} = \argmin\limits_{s' \in \mathcal{S}} d_{v,s'}$ denote the nearest center among a set of centers $\mathcal{S}$ and let $d_{v,\mathcal{S}} = d_{v,s_v^{\star}}$ be the corresponding minimum distance. It's easy to verify that $U(d_{v,\mathcal{S}}) = \min\limits_{s' \in \mathcal{S}} U(d_{v,s'})$ and $L(d_{v,\mathcal{S}}) = \min\limits_{s' \in \mathcal{S}} L(d_{v,s'})$ serve as the UCB and LCB respectively for $d_{v,\mathcal{S}}$.  

 The DS-UCB scheme, formally stated in Algorithm \ref{algo:dsucb}, works as follows and aims to emulate the stages of the greedy algorithm. Recall that the greedy algorithm chooses an arbitrary point $s_1$ as the first center and Algorithm \ref{algo:dsucb} does the same in line \ref{dsucb:randomfirst}. In each of the following stages, the scheme proceeds in rounds. In every round, to decide which arm $a_{v,s}$ to pull, we have to choose two indices $v$ and $s$.  From \eqref{Eqn:MABReward}, we have a maximization of $d_{v, \mathcal{S}}$ over the set of vertices $\mathcal{V}\backslash\mathcal{S}$, and so as the UCB algorithm from the MAB framework suggests, line \ref{dsucb:choosev} chooses the point $v$ with the highest UCB of $d_{v,\mathcal{S}}$ i.e., $U(d_{v,\mathcal{S}})$. Once the index $v$ is chosen, again from \eqref{Eqn:MABReward} we have a minimization of $d_{v,s}$ over the current set of centers $\mathcal{S}$ and line \ref{dsucb:chooses} chooses the center $s$ with the lowest LCB of $d_{v,s}$ i.e., $L(d_{v,s})$. 

 Having found the indices $v$, $s$ according to the UCB-LCB Sampling rule described above, we pull the arm $a_{v,s}$ to improve the estimate and confidence intervals for the distance $d_{v,s}$. In order to avoid too many pulls of any individual arm, we put a hard cap on the total number of times we sample any arm and on reaching the limit, explicitly calculate the corresponding distance by querying over all the dimensions one by one and updating the confidence intervals appropriately as in line \ref{dsucb:maxpulls}. Note that this requires at most an additional $m$ queries. 

 At the end of every round, we update the point $v^L$  with the maximum LCB of the distance to the current set of centers, i.e.,
 $v^L = \underset{v' \in \mathcal{V}}{\arg\max} \mbox{ } L(d_{v', \mathcal{S}})$, and we check if the break condition has been reached in line \ref{dsucb:termination}. We add the point $v^L$ as the new center to the current set of centers $\mathcal{S}$ and proceed to the next stage if 
\begin{align}\label{eqn:terminationcondition}
    L(d_{v^L,\mathcal{S}}) > \underset{v'\in \mathcal{V}\backslash \{v^L\}}{\max} \mbox{ } U(d_{v',\mathcal{S}}),
\end{align}
i.e., we are confident that the distance from $v^L$ to $\mathcal{S}$ is greater than that for any other point. We repeat the above process till we have identified $k$ centers. 

 \textit{Remark.} From a computational complexity point of view, in each round, line \ref{dsucb:choosev} needs to find a maximum over $O(n)$ points in each round. Hence, a naive implementation will require $O(nk)$ comparisons across the $k$ stages. By storing the values of $U(d_{v,\mathcal{S}}), L(d_{v,\mathcal{S}}) \mbox{ } \forall v \in \mathcal{V}$ in a heap data structure (\cite{CLRS}), the maxima and minima can be found in constant time while the complexity of maintaining the heap is $O(\log(n))$. Thus, the computational complexity reduces from $O(nk)$ to $O(k\log n)$.

\textit{Remark.} Note that DS-UCB as described above does not use the triangle inequality property of the normalized squared distance metric and thus can even be applied when the distance function doesn't satisfy that property. Also, we can use techniques similar to those described in \cite{MasonAndTripathy} to use the triangle inequality relations for tightening the confidence intervals and thus further reducing the number of queries required.
\begin{algorithm}
\small
\caption{DS-UCB}
\label{algo:dsucb}
\begin{algorithmic}[1]
\STATE Input: Confidence parameter $\delta$; Parameter $C_\alpha$
\STATE $\mathcal{S} = \{s_1\}$, $s_1 = x_i$, $i \sim \mbox{Unif}[1,n]$, \label{dsucb:randomfirst}, $\delta' = \delta/n^2$
\STATE MaxPulls$ = m$, $\hat{d}_{v,s} = 0 \mbox{ } \forall v \in \mathcal{V}, \forall s \in \mathcal{S}$
\WHILE{$|\mathcal{S}|$ $< k$}
\STATE  Query oracle $\mathcal{O}({v',s_{|\mathcal{S}|}},j)$ $\forall v'$ once, $j \sim \mbox{Unif}[1,m]$  
\STATE Update $\hat d_{v,s}$, $U(d_{v,s})$, $L(d_{v,s})$ 
\WHILE{$L(d_{v^L,\mathcal{S}}) \leq$ $\underset{v'\in \mathcal{V}\backslash \{v^L\}}{\max}$ $U(d_{v',\mathcal{S}})$ \label{dsucb:termination}} 
\STATE  $v \leftarrow \underset{v' \in \mathcal{V}}{\arg\max}$ $U(d_{v',\mathcal{S}})$ \label{dsucb:choosev}
\STATE $s \leftarrow \underset{s' \in \mathcal{S}}{\arg\min}$  $L(d_{v,s'})$ \label{dsucb:chooses}
\IF{$t_{v,s}$ $<$ MaxPulls }
\STATE Query oracle $\mathcal{O}({v,s})$
\STATE Update $\hat d_{v,s}$, $U(d_{v,s})$, $L(d_{v,s})$ 
\ELSIF{$t_{v,s} ==$ MaxPulls}
\STATE $\hat d_{v,s} = U(d_{v,s}) = L(d_{v,s}) \leftarrow d_{v,s}$ \label{dsucb:maxpulls}
\ENDIF
\STATE Update $L(d_{v,\mathcal{S}})$, $U(d_{v,\mathcal{S}})$, $v^{L}$, $t_{v,s} \leftarrow t_{v,s} + 1$
\ENDWHILE
\STATE $\mathcal{S} \leftarrow \mathcal{S} \cup \{v^L\}$
\ENDWHILE
\STATE \textbf{Return} $\mathcal{S}$
\end{algorithmic}
\end{algorithm}

\subsubsection{DS-UCB: Correctness and Query Complexity}
 In this section we argue the correctness of DS-UCB and also present an upper bound of its query complexity. We start by introducing a key lemma and some notation.

 The following lemma argues that with probability at least $1-\delta$,  the confidence intervals for all the distances hold true during the entire course of the scheme. \\

\textbf{Lemma 1.} (\cite{KaufmannLIL}, Lemma 7) \textit{For any $\delta \in (0,1)$, the following event occurs with probability at least $(1-\delta)$:}
\begin{align}\label{eqn:Ealpha}
   \!\!\! \mathcal{E}_{\alpha} := \left\{|\hat{d}_{v,s}(t)-d_{v,s}| \leq \alpha_{v,s}, \forall v \in \mathcal{V}, \forall s \in \mathcal{S}, \forall t  \right\}
\end{align}
\textit{where $\alpha_{v,s} = \alpha_{v,s}(t, \delta') =  \sqrt{\frac{2\beta(t,\delta/n^2)}{t}}$, $\beta(t,\delta') = 2\log(125\log(1.12t)/\delta')$} and {$\delta' = \delta/n^2$}.

 Say a set of $p$ centers $\mathcal{S}_p$ has been identified and recall that $d_{v, \mathcal{S}_p}$ represents the distance of point $v$ to the nearest center in $\mathcal{S}_p$. Furthermore, let the bottleneck distance $d_{\mathcal{V},\mathcal{S}_p}$ be defined as
\begin{equation}
\label{Eqn:BD}
 v^\star_p = \underset{v' \in \mathcal{V}}{\arg\max} \mbox{ } d_{v', \mathcal{S}_p}; \mbox{ } d_{\mathcal{V}, \mathcal{S}_p} = d_{v^\star_p,\mathcal{S}_p}.
\end{equation}
Note that the naive greedy algorithm would select the node $v^\star_p$ as the new center at the end of this stage. Our claim is that if the event $\mathcal{E}_{\alpha}$ occurs, then the termination condition in \eqref{eqn:terminationcondition} can only be triggered when $v^L = v^\star_p$. Assume to the contrary that $v^L \neq v^\star_p$ satisfies \eqref{eqn:terminationcondition}, then we have 
$$
d_{v^L,\mathcal{S}_p} \ge L(d_{v^L,\mathcal{S}_p}) > U(d_{v^\star_p,\mathcal{S}_p}) \ge d_{v^\star_p,\mathcal{S}_p}
$$
which is a contradiction by the definition of $v^\star_p$ in \eqref{Eqn:BD}. Thus, in each stage of the scheme, the center selected by DS-UCB is the same as the one that the naive greedy algorithm would have chosen. 

Having argued that DS-UCB mimics the output of the naive greedy algorithm, we will now present an upper bound on its query complexity. For each center $s_i$ and $p \in \{i,i+1,\ldots,k\}$, define $m^p _{v,s_i}$ as
\begin{equation}\label{eqn:mdef}
    m^p_{v,s_i} = \max\{d_{v,s_i} - d_{v,\mathcal{S}_p}, \ d_{\mathcal{V},\mathcal{S}_p} - d_{v,s_i}\}
\end{equation}
where recall that $d_{v,\mathcal{S}_p}$ denotes the minimum distance from $v$ to a center in $\mathcal{S}_p$ and $d_{\mathcal{V},\mathcal{S}_p}$ is defined in \eqref{Eqn:BD}. The above term encapsulates how easy or hard it is to distinguish an arm $a_{v,s_i}$ from the minimum-distance arm connecting $v$ to $\mathcal{S}_p$ and also from the overall optimal arm corresponding to the bottleneck distance, and appears in our upper bound on the query complexity of DS-UCB.

\begin{theorem}
With probability at least $(1-\delta)$, DS-UCB returns the set of centers whose bottleneck distance is optimal up to a factor of 2, in $T$ queries which can be upper bounded as
\begin{align*} 
    & T \le \sum\limits_{v \in \mathcal{V}\backslash\mathcal{S}} \sum\limits^k_{i=1} \max\limits_{p=\{i,., k-1\}} \frac{c\log(\frac{n^2}{\delta})\log[2\log(\frac{2}{m^p_{v,s_i}})]}{(m^p_{v,s_i})^2} \wedge 2m   \\
    & + \sum\limits^k_{i=1} \sum\limits^{i-1}_{j=1} \max\limits_{p=\{j, \dots, i-1\}} \frac{c\log(\frac{n^2}{\delta})\log[2\log(\frac{2}{m^p_{s_i,s_j}})]}{(m^p_{s_i,s_j})^2} \wedge 2m .
\end{align*}
{where $a \wedge b = min\{a,b\}$.}
\label{Thm:DSUCB-UB}
\end{theorem}
We relegate the proof of the above theorem to Appendix~\ref{Sec:ProofThm1}. The above bound is instance-dependent and incorporates the hardness of the dataset via the $\{m^{p}_{v,s}\}$ terms which encode how easy or hard it is to identify the optimal node at each stage of the scheme. 
Note that the bound evaluates to $O(nmk)$, which is the query complexity of the naive greedy algorithm, if the minimum of all the individual terms inside the summations is $2m$. However, the query complexity of DS-UCB can indeed be much smaller if the $\{m^{p}_{v,s}\}$ terms are significant. In particular, if the dataset is such that the $\{m^{p}_{v,s}\}$ terms are significant. For example, we randomly chose $n=10$ nodes each from the TinyImageNet \cite{tinyimagenet} and Greenhouse gas sensor \cite{gassensors} datasets and ran Stage $1$ of the DS-UCB algorithm. Over several rounds, we found $m^1_{v,s}$ to range from $.04 - .14$ and $.16 - .57$ for the two datasets respectively and as expected, we found the number of queries to be much larger for the TinyImageNet dataset ($\sim 25k$ vs $\sim 1k$). 
%

\subsection{DS-TS}
\label{Sec:DS-TS}
 In this section, we make an observation from the analysis of the query complexity of the DS-UCB algorithm. Using this observation, we propose an improved version of the algorithm called the DS-TS algorithm.

 Recall from the description of the DS-UCB scheme, that for a given node $v$, the choice of the arm $a_{v,s}$ to be pulled in any round is done by LCB-sampling (see line \ref{dsucb:chooses} in Algorithm~\ref{algo:dsucb}) which selects the arm among $\{a_{v, s_i}\}$ with the lowest LCB in that round. This sampling rule agrees\footnote{We use the LCB instead of the UCB since the objective function is the minimum expected reward.} with the well-accepted UCB principle from the MAB literature \cite{Jamieson} for identifying the optimal arm. However note that in any stage $p$ of the scheme, for all the points $v$, except the optimal point $v^\star_p$, we do not really need to identify the arm which minimizes the distance to $\mathcal{S}_p$. We only need to identify if there exists at least one arm among $\{a_{v, s}\}$ for which the expected reward is lower than this threshold $d_{v^\star_p, \mathcal{S}_p}$. This variant of the problem is posed in the MAB literature as identifying whether there exists an arm whose expected reward is less than some given threshold, and has been recently studied in \cite{KaufmannTMS}, \cite{juneja2019sample}. In particular, \cite{KaufmannTMS} argued that for a problem instance where there is indeed an arm with its expected reward below the threshold, Thompson Sampling (TS) \cite{RussoTS} performs much better than LCB-sampling and that the fraction of times the optimal arm (one corresponding to $d_{v, \mathcal{S}_p})$) is pulled to the total number of pulls converges almost surely to 1 with time. On the other hand, from the expression in Theorem~\ref{Thm:DSUCB-UB}, we can see that under LCB-sampling, for any sub-optimal node $v$ there is a contribution from all arms $\{a_{v,s}\}$ to the query complexity. This suggests that using TS instead of LCB-sampling can provide significant benefits in terms of the query complexity. One distinction to note in the problem here from the formulations in \cite{KaufmannTMS}, \cite{juneja2019sample} is that the threshold $d_{v^\star_p, \mathcal{S}_p}$ is in fact unknown (although the sampling rule doesn't need that information, the stopping rule is based on that knowledge).  \\\\
 With the above intuition we propose the DS-TS algorithm \ref{algo:dsts}. The algorithm works in stages similar to the DS-UCB algorithm and in each round, the indices $v,s$ need to be chosen to decide which arm $a_{v,s}$ to pull (see lines \ref{dsucb:choosev} and \ref{dsucb:chooses} in Algorithm~\ref{algo:dsucb}). The choice of $v$ is made as before, selecting the point $v \in \mathcal{V}$ with the highest upper confidence bound of the estimate of $d_{v,\mathcal{S}}$ i.e, $U(d_{v,\mathcal{S}})$. However unlike DS-UCB, for selecting $s$ we now use Thompson Sampling instead of LCB-sampling. \\\\
 Finally, note that unlike the other points, for the optimal point $v^\star_p$ we indeed need a good estimate for the optimal distance $d_{v^\star_p, \mathcal{S}_p}$ to the current set of centers since that is the unknown threshold in this case. This is akin to best arm identification and TS is known to be inefficient for this purpose \cite{russo2016simple},  since it doesn't explore enough resulting in poor concentration bounds for the sub-optimal arms which in turn leads to issues in termination (\ref{eqn:terminationcondition}). To deal with this, instead of a pure\footnote{There are variants of TS which work better for pure exploration \cite{shang2020fixed} and we did try some of them like T3C and D-Tracking, but found the mixed TS-LCB Sampling strategy to perform the best in our (admittedly limited) experiments.} TS strategy, we propose to use a mixture of TS and LCB Sampling to choose the index $s$ which in turn decides which arm to pull in each round. We introduce a factor $z$, that controls the mixture as TS for $z$ fraction of queries and LCB Sampling for the remaining $(1-z)$ fraction.\\\\
 For the TS based algorithm, we model the reward of an arm to be drawn from a Bernoulli distribution with parameter $\mu_{v,s}$. We choose the prior of the parameter to be a Beta distribution,
\begin{equation}\label{eqn:betaprior}
    \pi_{v,s} = \beta(S,F) \mbox{; } S=1, F=1 \mbox{ } \forall v \in \mathcal{V}, \forall s \in \mathcal{S}.    
\end{equation}
Since the oracle doesn't, in reality return $\{0,1\}$, we use the equivalence proposed by \cite{Agrawal} \footnote{Alternately, equivalence proposed in \cite{Riou} could also be used with degeneration of the arm to a Multinomial distribution and Dirichlet distribution as prior}. When an arm $a_{v,s}$ is pulled, the reward $r_{v,s}$ is considered to be the result of a Bernoulli trial with probability of success as the value returned by the oracle i.e., $r_{v,s}=$ Ber$(\mathcal{O}(v,s,j))$ where $j$ is a randomly chosen dimension. Given the reward $r_{v,s}$, the posterior distribution for $\mu_{v,s}$ is given by 
\begin{align}\label{eqn:posterior}
    \beta_{v,s}(& S(t_{v,s}+1), F(t_{v,s}+1)) = \nonumber\\ & \beta_{v,s}(S(t_{v,s}) + r_{v,s}, F(t_{v,s}) + (1-r_{v,s})).
\end{align}
Here, we use the confidence interval proposed in the KL Racing algorithm, \cite{KaufmannCI} as 
\begin{gather}
U(d_{v,s}) \leftarrow \max\left\{q:t_{v,s}kl(\hat{d}_{v,s},q) \leq \log\left(\frac{k_1t_{v,s}^\alpha}{\delta'}\right)\right\} \nonumber\\
L(d_{v,s}) \leftarrow \min\left\{q:t_{v,s}kl(\hat{d}_{v,s},q) \leq \log\left(\frac{k_1t_{v,s}^\alpha}{\delta'}\right)\right\} \nonumber\\
\alpha > 1; k_1 > 1 + \frac{1}{\alpha - 1} \label{eqn:klracing}
\end{gather}
where $kl(a,b)$ represents the Kullback-Leibler Divergence between distributions with mean parameters $a, b$. 

As done before, we cap the number of pulls of any individual arm and consider it to be out of contention for future rounds. The exit criterion for each phase is the same as DS-UCB and the algorithm terminates once $k$ centers have been identified. We compare the performance of DS-TS and DS-UCB via extensive numerical evaluations in Section~\ref{Sec:Experiments} and as our intuition suggested, the latter can provide significant benefits in terms of query complexity.

\begin{algorithm}
\small
\caption{DS-TS}
\label{algo:dsts}
\begin{algorithmic}[1]
\STATE Input: Confidence parameter $\delta$; Parameters { $\alpha, k_1, z$} 
\STATE $\mathcal{S} = \{s_1\}$, $s_1 = x_i$, $i \sim \mbox{Unif}[1,n]$
\STATE MaxPulls$ = m$, $\hat{d}_{v,s} = 0 \mbox{ } \forall v \in \mathcal{V}, \forall s \in \mathcal{S}$, $\delta' = \delta/n^2$
\STATE Choose a prior $\pi_{v,s}$ by (\ref{eqn:betaprior}) \label{dsts:chooseprior}
\WHILE{$|\mathcal{S}|$ $< k$}
\STATE Query oracle $O({v',s_{|\mathcal{S}|}},j) \mbox{ } \forall v'$ once, $j \sim \mbox{Unif}[1,m]$ 
\STATE Update $\hat d_{v,s}$, $U(d_{v,s})$, $L(d_{v,s})$ by (\ref{eqn:klracing})
\STATE $\mathcal{C} \leftarrow \{a_{v,s} | v \in \mathcal{V}, s \in \mathcal{S}\}$
\WHILE{$L(d_{v^L,\mathcal{S}}) \leq$ $\underset{v'\in \mathcal{V}\backslash \{v^L\}}{\max}$ $U(d_{v',\mathcal{S}})$}
\STATE  $v \leftarrow \underset{v' \in \mathcal{V}}{\arg\max}$ $U(d_{v',\mathcal{S}})$ \label{dsts:choosev}
\STATE $z_t = Ber(z)$; $\Theta_{v,s'} \sim \pi^{t-1}_{v,s'}$ $ \forall a_{v,s'}\in \mathcal{C}$ \label{dsts:tssample}
\STATE $s \leftarrow \underset{s' \in \mathcal{S}}{\arg\min}$ $z_t\{\Theta_{v,s'}\} + (1-z_t)\{L(v,s')\}$ \label{dsts:chooses}
\IF{$t_{v,s}$ $<$ MaxPulls }
\STATE Query oracle $\mathcal{O}(v,s,j)$, $j \sim \mbox{Unif}[1,m]$ 
\STATE Update $\hat d_{v,s}$, $U(d_{v,s})$, $L(d_{v,s})$ by (\ref{eqn:klracing})
\STATE Update posterior $\pi^t_{v,s}$ by (\ref{eqn:posterior})
\ELSIF{$t_{v,s} ==$ MaxPulls} 
\STATE $\hat d_{v,s} = U(d_{v,s}) = L(d_{v,s}) \leftarrow d_{v,s}$
\STATE $\mathcal{C} \leftarrow \mathcal{C}\backslash\{a_{v,s}\}$
\ENDIF
\STATE Update $L(d_{v,\mathcal{S}})$, $U(d_{v,\mathcal{S}})$, $v^{L}$, $t_{v,s} \leftarrow t_{v,s} + 1$
\ENDWHILE
\STATE $\mathcal{S} \leftarrow \mathcal{S} \cup \{v^L\}$
\ENDWHILE
\STATE \textbf{Return} $\mathcal{S}$
\end{algorithmic}
\end{algorithm}

\section{Noisy Distance Sampling Model} \label{Sec:NS}
 In this model (\ref{eqn:NSModel}), the oracle response to a query $(v, s)$ is the true distance between the pair of points corrupted by zero mean, variance $\sigma^2_{noise}$ Gaussian noise. While the specifics differ from the Dimension Sampling model, very similar ideas apply. In particular, the DS-UCB algorithm and its query complexity analysis in Theorem~\ref{Thm:DSUCB-UB} applies almost exactly under the Noisy Distance Sampling model, with the only change being a suitable choice of the confidence interval. For example, either Hoeffding's inequality \cite{MasonAndTripathy} or Kullback-Leibler divergence based confidence intervals \cite{KaufmannCI} hold true in this model. 

\subsection{NS-TS}
In this section, we present a Thompson Sampling based scheme NS-TS (Algorithm \ref{algo:nsts}) for this model, along similar lines as the DS-TS scheme. Similar to DS-TS, we choose the arm $a_{v,s}$ to pull in each round by selecting the index $v$ according to a UCB-based rule and the index $s$ by using a mixture of TS and LCB Sampling. For the TS part, we model the reward of an arm to be drawn from a Gaussian distribution $\mathcal{N}(\mu_{v,s}, \sigma^2_{noise})$, with the prior for $\mu_{v,s}$ also being chosen to be the Gaussian distribution $\mathcal{N}(0,1/2)$. 

We estimate the distance $d_{v,s}$ as
\begin{align}\label{eqn:nstsdistanceupdate}
    \hat{d}_{v,s}(t_{v,s}+1) = \frac{t_{v,s}}{t_{v,s}+1}\hat{d}_{v,s}(t_{v,s}) + \frac{1}{t_{v,s}+1}r_{v,s}(t_{v,s}).
\end{align}
The prior is chosen to be a Gaussian distribution as
\begin{equation} \label{eqn:gaussianprior} 
    \pi_{v,s} = \mathcal{N}(\mu, \sigma^2) \mbox{ with } \mu = 0, \sigma^2 = 0.5 \mbox{ } \forall v \in \mathcal{V}, \forall s \in \mathcal{S}. 
 \end{equation}

\noindent In line \ref{nsts:posteriorupdate}, the posterior is updated as 
\begin{gather}
    \pi_{v,s}^{t+1} = \mathcal{N}(\mu_{t+1}, \sigma^2_{t+1} ) \nonumber\\
    a = \frac{1}{\sigma^2_{t}}, b = \frac{t+1}{\sigma^2_{noise}} \nonumber\\
    \mu_{t+1} = \frac{a\mu_{t} + b\hat{d}_{v,s}(t+1)}{a+b}, \sigma^2_{t+1} = \frac{1}{a+b}. \label{eqn:gaussianposterior}
\end{gather}

\begin{algorithm}
\small
\caption{NS-TS}
\label{algo:nsts}
\begin{algorithmic}[1]
\STATE Input: Confidence parameter $\delta$; Parameters { $\alpha, k_1, z$}
\STATE $\mathcal{S} = \{s_1\}$, $s_1 = x_i$, $i \sim \mbox{Unif}[1,n]$
\STATE $\hat{d}_{v,s} = 0 \mbox{ } \forall v \in \mathcal{V}, \forall s \in \mathcal{S}$, $\delta'=\delta/n^2$
\STATE Choose a prior $\pi_{v,s}$ by (\ref{eqn:gaussianprior}) \label{nsts:chooseprior}
\WHILE{$|\mathcal{S}|$ $< k$}
\STATE Query oracle $\mathcal{O}({v',s_{|\mathcal{S}|}}) \mbox{ } \forall v'$ once 
\STATE Update $\hat d_{v,s}$, $U(d_{v,s})$, $L(d_{v,s})$ by (\ref{eqn:nstsdistanceupdate}) (\ref{eqn:klracing})
\WHILE{$L(d_{v^L,\mathcal{S}}) \leq$ $\underset{v'\in \mathcal{V}\backslash \{v^L\}}{\max}$ $U(d_{v',\mathcal{S}})$}
\STATE  $v_1 \leftarrow \underset{v' \in \mathcal{V}}{\arg\max}$ $\hat{d}_{v',\mathcal{S}}$, $v_2 \leftarrow \underset{v' \in \mathcal{V}\backslash \{v_1\}}{\arg\max}$ $U(d_{v',\cdot})$ \label{nsts:choosev}
\FOR{node $v \in \{v_1, v_2\}$}
\STATE $z_t = Ber(z)$; $\Theta_{v,s'} \sim \pi^{t-1}_{v,s'}$ $\forall a_{v,s'}$ \label{nsts:tssample}
\STATE $s \leftarrow \underset{s' \in \mathcal{S}}{\arg\min}$ $z_t\{\Theta_{v,s'}\} + (1-z_t)\{L(v,s')\}$ \label{nsts:chooses}
\STATE Query oracle $\mathcal{O}(v,s)$
\STATE Update $\hat d_{v,s}$, $U(d_{v,s})$, $L(d_{v,s})$ by (\ref{eqn:nstsdistanceupdate}) (\ref{eqn:klracing})
\STATE Update posterior $\pi^t_{v,s}$ by (\ref{eqn:gaussianposterior}) \label{nsts:posteriorupdate}
\STATE Update $L(d_{v,\mathcal{S}})$, $U(d_{v,\mathcal{S}})$, $v^{L}$, $t_{v,s} \leftarrow t_{v,s} + 1$
\ENDFOR
\ENDWHILE
\STATE $\mathcal{S} \leftarrow \mathcal{S} \cup \{v^L\}$
\ENDWHILE
\STATE \textbf{Return} $\mathcal{S}$
\end{algorithmic}
\end{algorithm}

\subsection{NS-TandS}

 In this section, we present another scheme for the NS model which is based on the Track-and-Stop strategy, originally proposed for best arm identification in the MAB framework \cite{kaufmann16a}. 

 Consider the $p$-th stage of the $k$-center problem, where we have found $p$ centers already and are looking for the $(p+1)^{th}$ center. Let $\vb*{d}^p \in \mathbb{R}^{(n-p) \times p}$ represent the distances between the remaining points and the current set of centers $\mathcal{S}_p$. Then the objective of the sub-problem for this stage can be stated as 
\begin{equation}\label{eqn:subproblem}
s_{p+1} = v^\star_p = \underset{v' \in \mathcal{V} \ \mathcal{S}_p}{\arg\max} \mbox{ } d_{v', \mathcal{S}_p} = \underset{v' \in \mathcal{V} \ \mathcal{S}_p}{\arg\max} \mbox{ } \min\limits_{s' \in \mathcal{S}_p} d_{v', s'}. 
\end{equation}
In this section, we first consider a closely related multi-armed bandit problem. We then propose the NS-TandS algorithm that utilizes a scheme inspired by ideas from the solution of the bandit problem in every stage of the $k$-center problem.

\subsubsection{Maximin Bandit}\label{Sec:maximinbandit}
Consider the following closely related multi-armed bandit problem presented in \cite{garivier2016maximin}. Let $\{a_{i,j}\}$ for $i \in \{1,\dots,a\}$ $j \in \{1,\dots,b\}$ denote a collection of arms spread across $a$ boxes, with $b$ arms each, where arm $a_{i,j}$ denotes the $j$-th arm in the $i$-th box. Further, let the random reward for each arm $a_{i,j}$ be  Gaussian distributed with mean $\mu_{i,j}$ and unit variance. Let $\vb*{\mu} \in \mathbb{R}^{a\times b}$ represent the vector of means of the arms. The objective of the problem is to find the maximin box $i^\star(\vb*{\mu}) = \argmax\limits_{i} \min\limits_j \mu_{i,j}$. Without loss of generality, we consider $\mu_{i,1} \leq \mu_{i,2} \leq \dots \leq \mu_{i,b}$ for each $i$ and $\mu_{1,1} \geq \mu_{2,1} \geq \dots \geq \mu_{a,1}$, so that the maximin box $i^\star = 1$.

The authors of \cite{garivier2016maximin} propose the D-Track-and-Stop algorithm for this problem which seeks to match the empirical fraction of pulls for the various arms to optimal weights derived as part of the solution to an optimization problem. This algorithm when coupled with an appropriate generalized-likelihood ratio based stopping rule has an asymptotic upper bound on the number of samples required which is within a constant factor of the information-theoretic lower bound, as stated in the theorem below.

\begin{theorem}[\cite{garivier2016maximin}, Proposition 6; \cite{kaufmann16a}, Proposition 13]\label{maximinupperbound}
With probability at least $(1-\delta)$ and for $\gamma \in [1,e/2]$, the D-Track-and-Stop algorithm returns the maximin box $i^\star(\vb*{\mu}) = 1$ in $\tau_\delta$ queries which can be asymptotically upper bounded as 
$$
\limsup\limits_{\delta \to 0} \frac{\tau_\delta}{\log(1/\delta)} \leq \gamma T^\star(\vb*{\mu})
$$
where
\begin{align}\color{red}
    & T^\star(\vb*{\mu})^{-1}  := \max\limits_{\vb*{\omega} \in \Tilde{\Omega}_{a,b}}  \min\limits_{i \neq 1,j} \mbox{ } \omega_{1,j}kl_g\left(\mu_{1,j}, \frac{\mu_{1,j}\omega_{1,j} + \mu_{i,1}\omega_{i,1}}{\omega_{1,j}+\omega_{i,1}}\right) \nonumber\\
    & {\color{white} ggggg} + \omega_{i,1}kl_g\left(\mu_{i,1}, \frac{\mu_{1,j}\omega_{1,j} + \mu_{i,1}\omega_{i,1}}{\omega_{1,j}+\omega_{i,1}}\right), \label{eqn:reducedTstar}\\
    & \Tilde{\Omega}_{a,b}  := \{\vb*{\omega} \in \mathbb{R}^{a \times b}_{+}: \sum\limits_i\sum\limits_j \omega_{i,j} = 1; \omega_{i,j}=0 \mbox{ } \forall i \geq 2, j \geq 2\}. \nonumber
\end{align}
\end{theorem}
{Define $\vb*{\omega}^\star \in \Tilde{\Omega}_{a,b}$ as the solution to the above optimization problem. Note that $kl_g(x,y) = (x - y)^2/2$ above denotes the Kullback-Leibler divergence between two Gaussian distributions with unit variance and means $x$ and $y$ respectively.} Furthermore, note from the definition of $\Tilde{\Omega}_{a,b}$ above that asymptotically, only arms $a_{1,j}$ belonging to the maximin box $i^\star = 1$ and the most competitive arm $a_{i,j}$ in each suboptimal box $j \neq 1$ are pulled.\\
{\textit{Remark.} Theorem 1 in \cite{kaufmann16a} provides the information theoretic lower bound for the maximin bandit as $\liminf_{\delta \to 0} \frac{\mathbb{E}[\tau_\delta]}{\log(1/\delta)} \geq T^\star(\vb*{\mu})$. The key difference between this and the lower bound stated in Theorem \ref{lowerboundtheorem} is that the expected rewards of arms in each box are independent of each other unlike in our problem when geometrical properties of points and spaces have to be respected.}

\subsubsection{NS-TandS: Algorithm}
From \eqref{eqn:subproblem}, we can see that in any stage of the greedy algorithm where $p$ centers have already been identified, if we view each pair of remaining vertex $i$ and existing center $j$ as an arm $a_{i,j}$ with mean reward $d_{i,j}$, the objective function for finding the next center is very similar to that in the maximin bandit problem with $n-p$ boxes, each containing $p$ arms. Recall that $\vb*{d}^p \in \mathbb{R}^{(n-p)\times p}$ represent the collection of these mean rewards $d_{i,j}$. The one subtle difference in the two formulations is that while in the bandit setting, the mean rewards $\{d_{i,j}\}$ are arbitrary and bear no relation to each other, in our $k$-center problem formulation, the mean rewards correspond to pairwise distances amongst a set of points and hence have to satisfy various constraints, for example the triangle inequality. However, for the purpose of designing a scheme NS-TandS for our problem, we ignore this difference and implement the D-Track-and-Stop (D-TaS) algorithm at each stage which tries to match the empirical proportions of queries of arms with $\vb*{w}^\star$ derived as a solution of the optimization problem (\ref{eqn:reducedTstar}). 

Formally stated in Algorithm~\ref{algo:nstrackstop}, NS-TandS works as described below and aims to emulate the stages of the greedy algorithm. The algorithm first chooses an arbitrary point $s_1$ as the first center in line 2. In each of the following stages, the algorithm proceeds in rounds choosing an arm $a_{v,s}$ to pull in each round in the following manner. After total $t$ rounds of oracle queries, let $t_{v,s}(t)$ be the number of times\footnote{Hereafter, $t_{v,s}(t)$ is abbreviated as $t_{v,s}$} arm $a_{v,s}$ is pulled and let $\vb*{\hat{d}}^p(t) \in \mathbb{R}^{(n-p)\times p }$ represent the estimates of the mean rewards $\{\hat{d}_{i,j}(t_{i,j})\}$. Define $\vb*{\omega}^\star$ as the optimal solution of (\ref{eqn:reducedTstar}). The objective of the algorithm in each stage is, now, to match the empirical proportion of times each arm is pulled with the fraction of pulls $\vb*{\omega}^\star(\vb*{d}^p)$. Since the true value of $\vb*{d}^p$ are unknown, the algorithm in every round first computes an estimate $\vb*{\omega}^\star(\vb*{\hat{d}}^p)$ in line~\ref{NSTaS:computew*} using the mirror gradient ascent algorithm (Appendix~C). In line~\ref{NSTaS:choosevs}, we choose the arm whose empirical fraction of pulls is the farthest away from the fraction needed. We also ensure that there is sufficient exploration of all arms, which is required for the correct output to be declared with high probability, by forcing all arms to be sampled at least {$\Omega(\sqrt{t})$} times in line~\ref{NSTaS:roott}. Further details on the  computation of $\vb*{\omega}^\star(\vb*{\hat{d}}^p)$ can be found in Appendix~C. 

After pulling an arm and improving our confidence of the estimates, in every round, the stopping condition checks if we are confident that the point whose estimate of the distance to its nearest center is the largest is indeed the next true  center in line~\ref{NSTaS:termination}. We use a stopping condition based on Chernoff's generalized likelihood ratio test \cite{Chernoff}. Define $\mu_{(i,j)}$ as the mean reward of arm $(i,j)$. For arms $(i,j), (i',j')$, consider the generalized likelihood ratio statistic
\begin{align*}
& Z_{(i,j)(i',j')}(t) := \\
& \log \frac{\max_{\mu'_{(i,j)} \geq \mu'_{(i',j')}} p_{\mu'_{(i,j)}}(\Bar{X}^{(i,j)}_{t_{(i,j)}})p_{\mu'_{(i',j')}}(\Bar{X}^{(i',j')}_{t_{(i',j')}})}{\max_{\mu'_{(i,j)} \leq \mu'_{(i',j')}} p_{\mu'_{(i,j)}}(\Bar{X}^{(i,j)}_{t_{(i,j)}})p_{\mu'_{(i',j')}}(\Bar{X}^{(i',j')}_{t_{(i',j')}})}
\end{align*}
where $\Bar{X}^{(i,j)}_{t_{(i,j)}}$ is a vector that contains the observations of arm $(i,j)$ available until time $t$, and where $p_{\mu}(Y_1, \dots, Y_n)$ is the likelihood of $n$ i.i.d. observations from a Gaussian distribution with mean $\mu$.
For our problem, the generalized likelihood ratio statistic for $\hat{d}_{i, j} \geq \hat{d}_{i',j'}$ can be restated as 
\begin{align*}
 Z_{(i,j)(i',j')}(t) := & t_{i,j}(t)kl_g(\hat{d}_{i,j}, \rho_{(i,j)(i',j')})\\
 & + t_{i',j'}(t)kl_g(\hat{d}_{i',j'}, \rho_{(i,j)(i',j')})
\end{align*}
where 
$$
\rho_{(i,j)(i',j')} = \frac{t_{i,j}(t)\hat{d}_{i,j}(t) + t_{i',j'}(t)\hat{d}_{i',j'}(t)}{t_{i,j}(t) + t_{i',j'}(t)}.
$$ 
and $Z_{(i',j')(i,j)}(t) = -Z_{(i,j)(i',j')}(t)$. If $\tau_\delta^p$ represents the total number of queries required for finding the $(p+1)^{th}$ center, the stopping condition for NS-TandS in line~\ref{NSTaS:termination} is given by 
\begin{align} \label{eqn:chernoffstopping}
    & \tau_\delta^p  = \inf \{t \in \mathbb{N} : \exists i \in \mathcal{V} : \forall i' \in \mathcal{V}\backslash\{i\}, \exists j' \in \mathcal{S}_p:\nonumber\\
    & {\color{white} afjhhhkgahdg} \forall j \in \mathcal{S}_p, Z_{(i,j)(i',j'))}(t) > \beta(t, \delta) \} \\
    & = \inf \{t \in \mathbb{N}: \max\limits_{i \in \mathcal{V}} \min\limits_{i' \in \mathcal{V}\backslash\{i\}} \max\limits_{j' \in \mathcal{S}_p} \min\limits_{j \in \mathcal{S}_p} Z_{(i,j)(i',j')}(t) > \beta(t,\delta)\} \nonumber
\end{align}
where $\beta(t,\delta)$ is an exploration rate to be tuned appropriately, which is typically chosen to be slightly larger than $\log(t/\delta)$. Defining $Z(t)$ as
\begin{align}
   Z(t) :=  \max\limits_{i \in \mathcal{V}} \min\limits_{i' \in \mathcal{V}\backslash\{i\}} \max\limits_{j' \in \mathcal{S}_p} \min\limits_{j \in \mathcal{S}_p} Z_{(i,j)(i',j')}(t),  \label{eqn:Zoft}
\end{align}
we have
$$
 \tau_\delta^p  = \inf \{t \in \mathbb{N} : Z(t) > \beta(t,\delta)\}. 
$$

\begin{algorithm}
\caption{NS-TandS}
\label{algo:nstrackstop}
\begin{algorithmic}[1]
\STATE Input: Confidence parameter $\delta$
\STATE $\mathcal{S} = \{s_1\}$, $s_1 = x_i$, $i \sim \mbox{Unif}[1,n]$
\STATE $\hat{d}_{v,s} = 0, t_{v,s} = 0 \mbox{ } \forall v \in \mathcal{V}, \forall s \in \mathcal{S}$; $\delta' = \delta/n^2$; $t=0$
\WHILE{$|\mathcal{S}| < k$}
\STATE Query oracle $\mathcal{O}({v',s_{|\mathcal{S}|}}) \mbox{ } \forall v'$ once, 
\STATE Update $\hat d_{v,s}$ by (\ref{eqn:nstsdistanceupdate})
\WHILE{$Z(t) \leq \beta(t,\delta')$ in (\ref{eqn:Zoft}) \label{NSTaS:termination}} 
\IF{$t_{v',s'} < \sqrt{t} - nk/2$}
\STATE $v,s \leftarrow \underset{v' \in \mathcal{V}, s' \in \mathcal{S}}{\arg\min}$ $t_{v',s'}$ \label{NSTaS:roott}
\ELSE
\STATE Compute $\vb*{\omega}^\star(\vb*{\hat{d}}^p)$ by (\ref{eqn:reducedTstar}) \label{NSTaS:computew*}
\STATE $v,s \leftarrow \underset{v' \in \mathcal{V}, s' \in \mathcal{S}}{\arg\max}$ $\omega^\star_{v',s'}(\vb*{\hat{d}}^p) - t_{v',s'}/t$  \label{NSTaS:choosevs}
\ENDIF
\STATE Query oracle $\mathcal{O}(v,s)$
\STATE Update $\hat d_{v,s}$ by (\ref{eqn:nstsdistanceupdate}), $t_{v,s} \leftarrow t_{v,s} + 1$, $t \leftarrow t + 1$
\ENDWHILE
\STATE $\mathcal{S} \leftarrow \mathcal{S} \cup \{\underset{v' \in \mathcal{V}}{\arg\max}$ $\min\limits_{s' \in \mathcal{S}}$ $\hat{d}_{v,s}\}$
\ENDWHILE
\STATE \textbf{Return} $\mathcal{S}$
\end{algorithmic}
\end{algorithm}

\subsubsection{NS-TandS: Correctness and Query Complexity}
Below, we state a result which argues the $\delta$-correctness of the NS-TandS scheme and also present an asymptotic upper bound on its query complexity, by considering each stage of finding a center sequentially. The proof of this theorem follows from \cite{kaufmann16a, garivier2016maximin} and is presented in Appendix~D for the sake of completeness.
\begin{theorem}\label{TaSupperboundtheorem} For any dataset with additive Gaussian noise, with probability at least $(1-\delta)$ and for $\gamma \in [1,e/2]$, NS-TandS returns the  set of centers whose bottleneck distance is optimal up to a factor of 2 in $\tau_\delta$ queries which can be upper bounded as 
$$
\limsup\limits_{\delta \to 0} \frac{\tau_\delta}{\log(1/\delta)} \leq \limsup\limits_{\delta \to 0} \sum\limits_{p=1}^{k-1} \frac{\tau_\delta^p}{\log(1/\delta)} \leq  \sum\limits_{p=1}^{k-1}  \gamma T^\star(\vb*{d}^p)
$$
where $T^\star(\vb*{d}^p)^{-1}$ is as defined in \eqref{eqn:reducedTstar}.
\end{theorem}

From the expression of $T^\star(\vb*{d}^p)^{-1}$ in \eqref{eqn:reducedTstar}, we know that $\omega^\star_{v,s} = 0$ $\forall v \in \mathcal{V}\backslash\{v^\star\},\forall s \in \mathcal{S}_p\backslash\{s^\star_v\}$. This means that only terms corresponding to all arms for the optimal point and the arms corresponding to the nearest center for the sub-optimal points are present in the above asymptotic upper bound expression. This matches our intuition from Theorem~\ref{lowerboundtheorem} and reaffirms our contention that queries from all arms are not necessary for identifying the $k$ centers.

\color{black}

\section{Experiments and Results}
\label{Sec:Experiments}
We evaluate the performance of our proposed schemes on the Tiny ImageNet (\cite{tinyimagenet}) and UT Zappos50K (\cite{zappos1}, \cite{zappos2}) datasets and compare their query complexity to that of the naive greedy algorithm as well as a Random Sampling algorithm which also progresses in stages but simply pulls an arm at random in each round. { In all our experiments except those reported in Figure \ref{fig:Calpha}, we observe that the set of centers returned by algorithms exactly matches the set of centers returned by the naive greedy algorithm which is used as the baseline.}

\subsection{DS-UCB}
 We use the dataset Tiny ImageNet for our experiments with each dimension normalized between [0,1]. The dataset has $64 \times 64 \times 3$ sized images which are reshaped into a $12288 \times 1$ vector. Further, we use the Euclidean distance metric and the iterated logarithm confidence interval of the form 
\begin{equation}
    \alpha(u) = \sqrt{\frac{C_\alpha \log(1+(1+\log(u))n^2/\delta)}{u}}
\end{equation}
where $C_{\alpha}$ is a hyperparameter. We use the naive implementation of the greedy algorithm as a reference and compare our results against it. 

 In each of the following set of experiments, the effect in change of one of the parameters is measured against the default setting of  $n=1000$, $k=10$, $m=12288$ in the Dimensional Sampling model. Each of the experiments is run 20 times for a fixed set of input images and a fixed first center and the average results are reported. The gains observed will depend on the specific instance of the bandit presented.

\begin{figure}
\centerline{\includegraphics[scale=0.45]{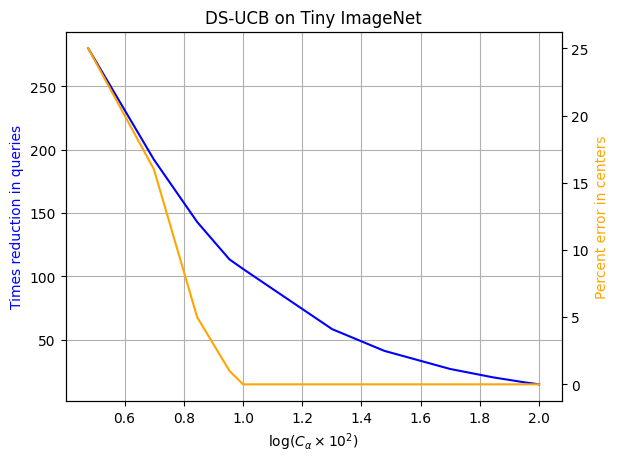}}
\caption{Queries and accuracy with changing $C_\alpha$ of DS-UCB on Tiny ImageNet dataset in comparison with naive greedy algorithm}
\label{fig:Calpha}
\end{figure}

 In Figure \ref{fig:Calpha}, { for $\delta=0.1$}, we observe that with decrease in $C_\alpha$, the width of the confidence interval decreases and hence the error rate increases while the number of queries decreases. Based on the trade off between error rate that we can tolerate against the performance, this is used to decide on the value of $C_\alpha$ and $C_{\alpha}=0.1$ used for the rest of the experiments.\footnote{This is along the lines of the experiment in LeJeune et al. (2019).} Table \ref{fig:delta} shows the increase in the number of queries as $\delta$ decreases and thus higher accuracy is demanded in the identification of the centers. We set $\delta$= 0.1 for the following experiments. Table \ref{fig:gain} presents a comparison of the performance of Naive algorithm, Random Sampling, and DS-UCB with varying $n$ and $k$. 

\begin{table}[]
\centering
\caption{Queries ($\times 10^4$) of DS-UCB with changing $\delta$ on Tiny ImageNet}
\begin{tabular}{|r|r|}
\hline
$\delta$ & DS-UCB  \\ \hline
0.3   & 107 \\
0.2   & 109 \\
0.1   & 116 \\
0.05  & 123 \\
0.01  & 137 \\
0.005 & 141 \\
0.001 & 157 \\
\hline
\end{tabular}
\label{fig:delta}
\end{table}

\begin{table}[]
\centering
\caption{Queries ($\times 10^5$) of Naive greedy vs Random Sampling vs DS-UCB on Tiny ImageNet }
\begin{tabular}{|r|r|r|r|r|}
\hline
$n$     & $k$  & \begin{tabular}[c]{@{}l@{}}Naive\\ Algorithm\end{tabular} & \begin{tabular}[c]{@{}l@{}}Random\\ Sampling\end{tabular} & DS-UCB  \\ \hline
100   & 10 & 123 & 191 & 11\\
500   & 10 & 614  & 864   & 8  \\
1000  & 10 & 1229 & 1536   & 12  \\
5000  & 10 & 6144   & 7913    & 28  \\
10000 & 10 & 12288 & 17994 & 47 \\ \hline
1000  & 3  & 369 & 254 & 2   \\
1000  & 5  & 614 & 591 & 4   \\
1000  & 7  & 860  & 883 & 6   \\
1000  & 10 & 1229 & 1536 & 12 \\
\hline
\end{tabular}
\label{fig:gain}
\end{table}

\subsection{DS-TS}
 We use the dataset Tiny ImageNet for our experiments with the distance in each dimension normalized to the interval [0,1]. For all the experiments, we use the Kullback-Leibler divergence based confidence intervals \cite{KaufmannCI} instead of those based on the law of iterated logarithm described before, since we find their performance to be better. With a fixed set of input images and a fixed first center, we run the DS-TS algorithm for $z=1$ (pure TS), $z=0.99$ (TS-LCB Sampling mixture) and $z=0$ (pure LCB Sampling). We also vary the values of $n$ and $k$ and see how it impacts the query complexity, where each experiment was run 20 times and the average results are reported. We observe in Table \ref{fig:DS-TS1} that TS-LCB Sampling mixture performs the best followed by pure TS and pure LCB Sampling. This matches our intuition from Section~\ref{Sec:DS-TS}. Also, Table \ref{fig:DS-TS} clearly demonstrates the superior performance of our scheme over the naive greedy and Random Sampling schemes. 

\begin{table}
\centering
\caption{Queries ($\times 10^4$) of DS-TS on Tiny Imagenet for different values of mixing parameter $z$}
\begin{tabular}{|r|r|r|r|r|}
\hline
$n$     & $k$  & \begin{tabular}[c]{@{}l@{}}DS-TS\\ $z = 1$\end{tabular} & \begin{tabular}[c]{@{}l@{}}DS-TS\\ $z = 0.99$\end{tabular} & \begin{tabular}[c]{@{}l@{}}DS-TS\\ $z = 0$\end{tabular} \\ \hline

100   & 10 & 271                                               & 268                                                  & 286                                               \\
500   & 10 & 305                                               & 287                                                  & 357                                               \\
1000  & 10 & 450                                               & 431                                                  & 588                                               \\
5000  & 10 & 1309                                              & 1305                                                 & 1869                                              \\
10000 & 10 & 2107                                              & 2113                                                 & 3057                                              \\ \hline
1000  & 3  & 75                                                & 75                                                   & 85                                                \\
1000  & 5  & 166                                               & 165                                                  & 195                                               \\
1000  & 7  & 271                                               & 267                                                  & 327                                               \\
1000  & 10 & 450                                               & 431                                                  & 588                                            \\ 
\hline
\end{tabular}
\label{fig:DS-TS1}
\end{table}

\begin{table}[]
\centering
\caption{Queries ($\times 10^5$) of Naive greedy vs Random Sampling vs DS-TS on Tiny Imagenet}
\begin{tabular}{|r|r|r|r|r|}
\hline
$n$     & $k$  & \begin{tabular}[c]{@{}l@{}}Naive\\ Algorithm\end{tabular} & \begin{tabular}[c]{@{}l@{}}Random \\ Sampling\end{tabular} & \begin{tabular}[c]{@{}l@{}}DS-TS\\ $z = 0.99$\end{tabular} \\ \hline

100   & 10 & 123                                                  & 189                                                   & 27                                                  \\
500   & 10 & 614                                                  & 892                                                   & 29                                                  \\
1000  & 10 & 1229                                                 & 1731                                                  & 43                                                  \\
5000  & 10 & 6144                                                 & 9856                                                  & 130                                                 \\
10000 & 10 & 12288                                                & 19790                                                 & 211                                                 \\ \hline
1000  & 3  & 369                                                  & 406                                                   & 7                                                   \\
1000  & 5  & 614                                                  & 518                                                   & 16                                                  \\
1000  & 7  & 860                                                  & 1305                                                  & 27                                                  \\
1000  & 10 & 1229                                                 & 1731                                                  & 43                           \\

\hline                      
\end{tabular}
\label{fig:DS-TS}
\end{table}

\subsection{NS-TS}
\subsubsection{Tiny ImageNet}
 For the Noisy Distance Sampling model, we again consider the Tiny Imagenet dataset and assume that each pairwise distance query is corrupted by zero mean, $0.01$ variance Gaussian noise. With a fixed set of input images and a fixed first center, we run the NS-TS algorithm\footnote{The Random Sampling algorithm takes many more queries and did not terminate after an extended period of time for many of our experiments.} for $z=0.9, 0.8$ (TS-LCB Sampling Mixture) and $z=0$ (Pure LCB Sampling). We take $n=1000$ and vary the value of $k$, where each experiment is run 10 times and the average results are reported in Table \ref{fig:NS-TS}. One thing to observe is that while there isn't a universally optimal value of the mixing parameter $z$, we found that higher values close to $0.9$ (more TS, less LCB) seem to perform better in most cases. For a noise variance of $0.001$, we compare the performance of NS-TS with NS-TandS in Table~\ref{fig:NS-TaS}. We find that there are several scenarios where NS-TandS performs better from NS-TS. However, we also observe the opposite behavior in some cases where the number of points is relatively high compared to the number of centers to be found, and we believe this is because the forced exploration of $O(\sqrt{t})$ queries for every arm in NS-TandS dominates the savings we get from optimal tracking of sampling frequencies. A more principled way of tuning the forced exploration is yet to be found for Track and Stop based algorithms  \cite{Lattimore}. \color{black}

\begin{table}[]
\centering
\caption{Queries ($\times 10^3$) of NS-TS on Tiny Imagenet for different values of mixing parameter $z$}
\begin{tabular}{|r|r|r|r|r|}
\hline
$n$    & $k$  & \begin{tabular}[c]{@{}l@{}}NS-TS \\ $z = 0.9$\end{tabular} & \begin{tabular}[c]{@{}l@{}}NS-TS \\ $z = 0.8$\end{tabular} & \begin{tabular}[c]{@{}l@{}}NS-TS \\ $z = 0$\end{tabular} \\ \hline
100  & 3  & 31                                                    & 27                                                    & 29                                                  \\
100  & 5  & 170                                                   & 132                                                   & 138                                                 \\
100  & 7  & 1120                                                  & 1088                                                  & 1098                                                \\
100  & 10 & 11326                                                 & 11142                                                 & 11137                                               \\ 
\hline
500  & 3  & 16                                                    & 16                                                    & 17                                                  \\
500  & 5  & 76                                                    & 80                                                    & 98                                                  \\
500  & 7  & 2147                                                  & 2277                                                  & 2637                                                \\
500  & 10 & 2447                                                  & 2573                                                  & 2956                                                \\ 
\hline
1000 & 3  & 31                                                    & 32                                                    & 34                                                  \\
1000 & 5  & 108                                                   & 107                                                   & 122                                                 \\
1000 & 7  & 297                                                   & 310                                                   & 387                                                 \\
1000 & 10 & 2070                                                  & 1636                                                  & 1651                                         \\

\hline     
\end{tabular}
\label{fig:NS-TS}
\end{table}

\begin{table}[]
    \centering
    \caption{Queries ($\times 10^2$) of NS-TandS vs NS-TS on Tiny Imagenet}
    \begin{tabular}{|c|c|c|c|c|}
    \hline
    $n$    & $k$  & \begin{tabular}[c]{@{}l@{}}NS-TS \\ $z = 0$\end{tabular} & \begin{tabular}[c]{@{}l@{}}NS-TS \\ $z = 0.8$\end{tabular} & \begin{tabular}[c]{@{}l@{}}NS-TandS \\ \end{tabular} \\ \hline
    15     & 3 & 137 & 137 & 85\\
    15     & 5 & 208 & 204 & 137\\
    15     & 10 & 5326 & 4800 & 2703\\ \hline
    25     & 3 & 32 & 30 & 30\\
    25     & 5 & 180 & 157 & 206\\
    25     & 10 & 3708 & 3692 & 1750\\ \hline
    50     & 3 & 29 & 25 & 71\\
    50     & 5 & 600 & 598 & 1085\\
    50     & 10 & 44151 & 35814 & 14741\\
    \hline
    \end{tabular}
    \label{fig:NS-TaS}
\end{table}

\subsubsection{UTZappos50k}
\begin{figure}
\centerline{\includegraphics[scale=0.12]{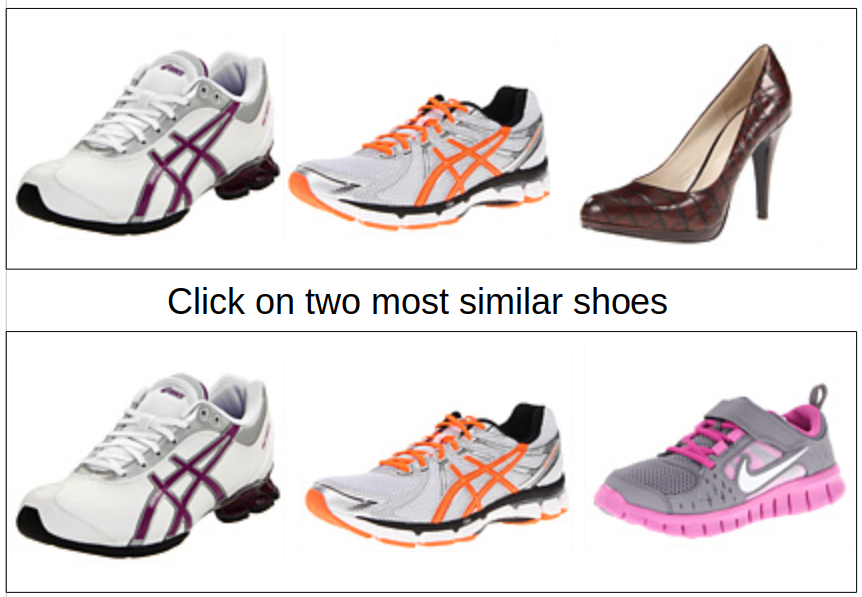}}
\caption{Two examples of zappos queries. Top one is an easy query and the bottom one is a hard query.}
\label{fig:zapposquery}
\end{figure}

 While the previous experiments used the normalized squared distance and artificially induced noise, in this experiment we consider the noise to be a result of human judgments and use an appropriate data-driven distance metric. \cite{Heim} considered a set of 85 images of shoes drawn from the UTZappos50K dataset \cite{zappos1}, \cite{zappos2} and gathered responses on the Amazon Mechanical Turk to queries of the form ``between images $i,j, \mbox{ and, } l$ which two are the most similar?" over all possible triplets to find which shoes are the most similar. For example, in the top image in Figure \ref{fig:zapposquery}, the second shoe is considered more similar to the first one that the third one. We can also see that the top image represents an easier query than the bottom one, which presents three very similar shoes. Based on these responses, a heuristic distance measure was proposed in \cite{MasonAndTripathy} which defined the distance between two images, $d_{i,j}$, as the fraction of times that the images $i,j$ are judged as being more similar to each other than a randomly chosen third image $l$. It can be stated mathematically as
\begin{align}
    d_{i,j} = 1 - \mathbb{P}(``i & \mbox{ more similar to }j\mbox{ than }l")\times \nonumber\\
    & \mathbb{P}(``j\mbox{ more similar to }i\mbox{ than }l") \nonumber
\end{align}
where $l \sim \mbox{Unif}(\mathcal{V}\backslash\{i,j\})$ and probabilities are the empirical probabilities of images in the dataset.

 Using $n=55$ shoes as the points and the above defined heuristic distance measure, we ran the NS-TS algorithm for identifying $k = 3,5,7,10$ centers (which will correspond to representative footwear), and compared its performance on this dataset to the Random Sampling scheme. For the NS-TS scheme, we consider the reward from a pull of arm $a_{v,s}$ as $r_{v,s} = \mbox{Ber}(d_{i,j})$. We consider a Beta prior $\beta(1,1)$. We report the results in Table \ref{fig:zapposgain} where we again observe that the number of queries required by the NS-TS algorithm is significantly lower than the Random Sampling policy. From Table~\ref{fig:zapposgain}, we also observe that NS-TandS performs the best among all the schemes. To save computational effort and improve the running time of the NS-TandS algorithm, we perform the sub-task of finding the optimal weights $\vb*{\omega}^\star(\hat{\vb*{d}}^p)$ by the mirror gradient ascent algorithm once every $100$ queries to the oracle. \color{black}

\begin{table}[]
\centering
\caption{Queries ($\times 10^4$) for UTZappos50k dataset}
\begin{tabular}{|r|r|r|r|r|r|r|}
\hline
$n$  & $k$  & \begin{tabular}[c]{@{}l@{}}Random\\ Sampling\end{tabular} & \begin{tabular}[c]{@{}l@{}}NS-TS\\ $z = 0.9$\end{tabular} & \begin{tabular}[c]{@{}l@{}}NS-TS\\ $z = 0.7$\end{tabular} & \begin{tabular}[c]{@{}l@{}}NS-TS\\ $z = 0$\end{tabular} & NS-TandS \\ \hline
55 & 3  & 3242                                                  & 182                                                 & 183                                                & 174 & 79                                              \\
55 & 5  & 6635                                                  & 269                                                 & 263                                                 & 259  & 247                                             \\
55 & 7  & 40878                                                 & 589                                                 & 593                                                 & 600 & 323                                               \\
55 & 10 & 41765                                                 & 724                                                 & 733                                                 & 732 & 603\\ 
 \hline
55 & 5  & 6635                                                  & 269                                                 & 263                                                 & 259 & 247                                               \\ 
45 & 5  & 34823                                                 & 606                                                 & 606                                                 & 614 & 195                                              \\
35 & 5  & 65649                                                 & 952                                                 & 957                                                 & 961 & 158                                              \\
25 & 5  & 1203                                                  & 95                                                  & 93                                                  & 92 & 67                                                \\
15 & 5  & 3943                                                  & 248                                                 & 248                                                 & 250 & 132\\
\hline
\end{tabular}
\label{fig:zapposgain}
\end{table}

\section{Conclusion}
\label{Sec:Conclusions}
In this work, we studied a variant of the canonical $k$-Center problem where only noisy/incomplete estimates of distances between pairs of vertices are available. We cast this into a multi-armed bandit framework where each arm of the bandit represented the distances between pairs of points. We designed sequential algorithms based on the principle of adaptive sampling and ideas from bandit literature such as UCB, Thompson Sampling and Track-and-Stop. In addition to being suitable for the setting of noisy information availability, these algorithms also provide computational savings over naive implementations in terms of the number of pairs of vertices sampled and the number of dimensions sampled between each pair. We established a fundamental lower bound on the query complexity of the $k$-Center problem for a restricted class of algorithms and datasets. We also proved instance-dependent upper bounds on the query complexities of a subset of the policies. Finally, we experimentally demonstrated the validity of our claims through numerical evaluations on two real-world datasets - Tiny ImageNet and UT Zappos50K. Proving a more general lower bound for this setting and designing algorithms to match this lower bound are open questions. 
\color{black}

\begin{appendices}
\section{Proof of Theorem~\ref{lowerboundtheorem}}
\label{lowerboundproof}

{\color{black} Recollect that we consider an algorithm $\mathcal{A}$ that mimics the greedy algorithm by proceeding in stages and iteratively building the solution set by choosing a new center at every stage $p = \{1, \dots, k-1\}$. Also recall the MAB mapping for the $k$-Center problem established in Section~\ref{section:problem} with each node corresponding to a box containing multiple arms, one for the distance to each of the current centers. We define the score associated with a box as the minimum expected reward of its arms. The goal in each stage then is the find the box with the highest score, which maps to the next picked center. To establish a lower bound on the query complexity, we proceed using the standard technique in the MAB literature which involves  identifying the minimum number of queries required to differentiate the current instance from an alternate instance which has a different best arm. Using this technique, for our problem, the minimum number of expected samples corresponds to the least number of samples required to distinguish the current problem instance from an alternate problem instance in which the set of centers and/or the order that they are identified in is different. For example, if the set of centers identified for the original instance is $(s_1, s_2, \dots, s_k)$, then instances that result in the set of centers being identified as either $(s_1, v \in \mathcal{V} \backslash \{s_2\}, s_3, \dots, s_k)$ or $(s_2, s_1, \dots, s_k)$ are considered as alternate instances.}

{\color{black} In the proof below, we first state a useful lemma which is useful in identifying the minimum number of expected samples from each arm that are required to distinguish between two problem instances with different reward distributions. We then construct alternate instances such that the optimal box is different from the original instance in at least one stage. Finally, using these constructed alternate instances, we derive a lower bound on the expected number of samples required from the boxes corresponding to the set of non-center points i.e. $v' \in \mathcal{V} \backslash \mathcal{S}_k$ and then the samples required from the boxes corresponding to the desired set of centers $\mathcal{S}_k$.}

{\color{black}
\textbf{Lemma 2.} (\cite{KaufmannLIL}, Lemma 1) \textit{Let $\nu, \nu'$ be two collections of $r$ probability distributions over $\mathbb{R}$, each representing the reward distributions for $r$ arms in a standard MAB framework. Then for any event $\mathcal{E}\in\mathcal{F}_{\xi}$ with stopping time $\xi$ and $P_{\nu}[\mathcal{E}] \in (0,1)$, we have }
\begin{equation}
    \sum\limits_{l=1}^r \mathbb{E}_{\nu}[T_{l}(\xi)]kl(\nu_{l}, \nu_{l}') \geq kl(P_{\nu}[\mathcal{E}], P_{\nu'}[\mathcal{E}]) 
    \label{eqn:changeofmeasure}
\end{equation}
\textit{where $T_{l}$ is the number of pulls of arm $l$ and} $kl$ \textit{is the KL divergence.}
}

{\color{black} Next, we adapt Lemma 2 to our problem. In Lemma 2, if $\nu, \nu'$ are selected such that the best box $(v^\star_p)_\nu$ under the original instance $\nu$ is different from the best box $(v^\star_p)_{\nu'}$ under the alternate instance $\nu'$  for at least one stage $p$, and the event $\mathcal{E}$ is chosen to be that the output of the algorithm matches the greedy algorithm, i.e. $(s_1, s_2, \dots, s_k)$, then our requirement is $P_\nu (\mathcal{E}) \geq 1-\delta$, $P_{\nu'}(\mathcal{E}) \leq \delta$. For this choice of $\mathcal{E}$, we have from \eqref{eqn:changeofmeasure} that 
\begin{align}\label{eqn:lowerboundbasic}
    \sum\limits_{v \in \mathcal{V}\backslash\mathcal{S}_p} \sum\limits_{s \in \mathcal{S}_p} \mathbb{E}_{\nu}[T_{v,s}(\xi)]kl(\nu_{v,s}, \nu_{v,s}') & \geq kl(\delta, 1-\delta) \nonumber\\
    & \geq \log(1/2.4\delta)
\end{align}
}
{\color{black} Recollect that for our setting, $\nu$ is our given distribution with the reward distribution corresponding to each arm $(i,j)$ being Ber($d_{i,j}$) since  the oracle returns $-1/2$ or $1/2$ as the answer to every query. Consider the stage $p \in \{1,2, \dots, k-1\}$ where $p$ centers have been identified and the next center is being sought. We now construct alternate instances.}

{\color{black} First, consider the boxes corresponding to the set of non-center points $v' \in \mathcal{V} \backslash \mathcal{S}_k$. To make $v'$ the optimal box at stage $p$ instead of the the desired center $s_{p+1}$, we have to make the expected value of the arms in the box $v'$ greater than score of box $s_{p+1}$, which is $d_{\mathcal{V}, \mathcal{S}_p}$. By moving $v'$ to near $s_{p+1}$ appropriately and keeping distributions corresponding to the rest of the boxes the same, we change the score of box $v'$ to $d_{\mathcal{V}, \mathcal{S}_p} + \epsilon$.\footnote{Note the such a move is possible in all cases except when $s_{p+1}$ is the circumcenter of the polygon formed by the existing $\mathcal{S}_p$, which is highly unlikely for high dimensional real datasets in which we are primarily interested in this work} Note that $\forall v \neq v'$ and any $s \in \mathcal{S}_p$, we have $kl(\nu_{v,s}, \nu'_{v,s}) = 0$  since the alternate instance $\nu'$ constructed keeps the distributions of arms in boxes other than $v'$ the same as the original instance $\nu$. Thus, we have from \eqref{eqn:lowerboundbasic} that 
\begin{align}
    & \sum\limits_{i=\{1,\dots,p\}} kl_b(d_{v',s_i}, d_{s_{p+1}, s_i}) \cdot \mathbb{E}_{\nu}[T_{v',s_i}(\xi)] \geq \log(1/2.4\delta) \nonumber\\
    \Longrightarrow & \max\limits_{i=\{1,\dots,p\}} kl_b(d_{v',s_i}, d_{s_{p+1}, s_i}) \cdot \sum\limits_{i=\{1,\dots,p\}} \mathbb{E}_{\nu}[T_{v',s_i}(\xi)] \nonumber\\
    &\hspace{2.4in}\geq \log(1/2.4\delta) \nonumber\\
    \Longrightarrow & \mathbb{E}_{\nu}[T^p_{v'}] \geq \frac{\log(1/2.4\delta)}{\max\limits_{i=\{1,\dots,p\}}kl_b(d_{v',s_i}, d_{s_{p+1}, s_i})} 
\end{align}
where we let $T^p_{v'}$ denote the total number of pulls from a box $v' \in \mathcal{V} \backslash (\mathcal{S}_k)_{\nu}$ at stage $p$ and $kl_b(x,y)$ denotes the KL divergence between two Bernoulli distributions with means $x$ and $y$.}

{\color{black} Taking the maximum over all possible stages $p \in \{1, \dots, k-1\}$ at which $v'$ can replace the desired center to create an alternate instance, we have
\begin{align}\label{eqn:lowerbounda}
    \mathbb{E}_{\nu}[T_{v'}] & \geq \max\limits_{p=\{1,\dots,k-1\}} \mathbb{E}_{\nu}[T^p_{v'}] \nonumber \\
    & \geq \max\limits_{p=\{1,\dots,k-1\}} \frac{\log(1/2.4\delta)}{\max\limits_{i=\{1,\dots,p\}}kl_b(d_{v',s_i}, d_{s_{p+1}, s_i})}.
\end{align}}

{\color{black} Now consider the box corresponding to the desired center found at stage $p$, $s_{p+1}$. To create an alternate instance by excluding this point from the set of centers, we decrease the score of this box to $d_{v'', \mathcal{S}_p}-\epsilon$, for some  $v'' \in \mathcal{V} \backslash \mathcal{S}_{p+1}$. This can be achieved by moving the point $s_{p+1}$ to near $v''$ appropriately. Thus, $v''$ replaces $s_{p+1}$ as the center at stage $p$. Taking a maximum over all possible $v''$ replacements for $s_{p+1}$ that create an alternate instance, we have from equation (\ref{eqn:lowerboundbasic}),
\begin{equation}\label{eqn:lowerboundb}
    \mathbb{E}_{\nu}[T_{v^\star_p}] \geq \max\limits_{v'' \in \mathcal{V} \backslash \mathcal{S}_{p+1}} \frac{\log(1/2.4\delta)}{\max\limits_{i=\{1,\dots,p\}}kl_b(d_{s_{p+1}, s_i}, d_{v'',s_i})}.
\end{equation}}

{\color{black} Combining equations (\ref{eqn:lowerbounda}) and (\ref{eqn:lowerboundb}), and accounting for arms that could be counted twice, we get the lower bound
\begin{align}
    \mathbb{E}[T] \geq \frac{1}{2} \left(\sum\limits_{v' \in \mathcal{V} \backslash \mathcal{S}_k} \mathbb{E}_\nu[T_{v'}] + \sum\limits_{p=1}^{k-1} \mathbb{E}_{\nu}[T_{v^\star_p}] \right).
\end{align}}
{\color{black}This matches the expression given in the statement of the theorem and thus completes the proof.}

{\color{black}We would like to remark that using $\ln x \leq x - 1$, we can bound the KL divergence between two Bernoulli distributions as $kl_b(a,b) \leq \frac{(a-b)^2}{b(1-b)}$. Hence equations (\ref{eqn:lowerbounda}) and (\ref{eqn:lowerboundb}) can be further bounded as 
\begin{gather}
    \mathbb{E}_{\nu}[T_{v'}]
    \geq \max\limits_{p=\{1,\dots,k-1\}} \min\limits_{i = \{1,\dots,p\}} \frac{\log(1/2.4\delta) d_{v^\star_p, s_i}(1-d_{v^\star_p, s_i}) }{(d_{v',s_i}- d_{v^\star_p, s_i})^2}, \nonumber \\
    \mathbb{E}_{\nu}[T_{v^\star_p}]
    \geq \max\limits_{v'' \in \mathcal{V} \backslash \mathcal{S}_{p+1}} \min\limits_{i = \{1,\dots,p\}} \frac{\log(1/2.4\delta) d_{v'',s_i}(1-d_{v'',s_i}) }{(d_{v^\star_p, s_i}- d_{v'',s_i})^2}. \nonumber
\end{gather}}

{\color{black} We observe that the form of these expressions is similar to the terms in the upper bound in Theorem \ref{Thm:DSUCB-UB}, with both showing inverse dependence on the square of the difference between the distance terms.}

\section{Proof of Theorem~\ref{Thm:DSUCB-UB}}
\label{Sec:ProofThm1}
 We state a fact below that will be useful in proving our upper bound on the query complexity of the DS-UCB scheme. 

\textbf{Fact.} (\cite{LeJeune}, Fact 6). \textit{For $\delta'>0$, if $T$ is the smallest integer $u$ satisfying the bound $\alpha(u, \delta') \le \Delta/8$, then for c sufficiently large,} 
\begin{equation}\label{eqn:fact}
    T \le c\log\left(\frac{1}{\delta'}\right) \frac{\log(2\log(2/\Delta))}{\Delta^2}.
\end{equation}
For each stage, we consider the number of times any arm $a_{v, s_i}$ is pulled, where the current set of centers is $\mathcal{S}_p$. Taking the worst across all stages of those bounds gives us an upper bound on the number of pulls of that arm across the run of the algorithm.

 If $\mathcal{E}_{\alpha}$ (\ref{eqn:Ealpha}) occurs, arm $a_{v, s_i}$ will not be pulled anymore when either of the following happens.\\
\textbf{(A)} In the minimization process, for a given $v$, the arm pulled corresponds to the one with the lowest LCB $L(d_{v,s})$ and since the minimum LCB will certainly be smaller than $d_{v, \mathcal{S}_p}$, a sufficient condition for the arm to not be pulled any longer is
\begin{gather}
   L(d_{v, s_i}) \ge d_{v,\mathcal{S}_p}.\\
   d_{v,s_i} - 2\alpha_{v,s_i} \ge d_{v,\mathcal{S}_p} \implies \alpha_{v,s_i} \le \frac{d_{v,s_i} - d_{v,\mathcal{S}_p}}{2}\nonumber
\end{gather}
From (\ref{eqn:fact}), we have an upper bound on the number of times arm $a_{v,s_i}$ is pulled as\\
\begin{align}\label{eqn:blah1}
      T^{A,p}_{v,s_i} \le \frac{c\log(\frac{n^2}{\delta})\log[2\log(\frac{2}{d_{v,s_i}-d_{v,\mathcal{S}_p}})]}{(d_{v,s_i}-d_{v,\mathcal{S}_p})^2} \wedge 2m.
\end{align}
\textbf{(B)} In the maximization process, the algorithm chooses a point $v$ with the highest UCB $U(d_{v, \mathcal{S}_p})$ and since the maximum UCB will certainly be greater than $d_{\mathcal{V},\mathcal{S}_p}$, a sufficient condition for point $v$ to not be chosen any longer is
\begin{align}\label{eqn:tsrep}
    U(d_{v,s}) \le d_{\mathcal{V},\mathcal{S}_p} \mbox{ for any } s \in \mathcal{S}_p.
\end{align}
A sufficient condition for this criterion is
\begin{gather}
      U(d_{v,s_i}) \leq d_{\mathcal{V}, \mathcal{S}_p}.\\
      d_{v,s_i} + 2\alpha_{v,s_i} \le d_{\mathcal{V},\mathcal{S}_p} \implies \alpha_{v,s_i} \le \frac{d_{\mathcal{V},\mathcal{S}_p} - d_{v,s_i}}{2} \nonumber
\end{gather}
From (\ref{eqn:fact}), again, we have,\\
\begin{equation}\label{eqn:blah2}
     T^{B,p}_{v,s_i} \le \frac{c\log(\frac{n^2}{\delta})\log[2\log(\frac{2}{d_{\mathcal{V},\mathcal{S}_p}-d_{v,s_i}})]}{(d_{\mathcal{V},\mathcal{S}_p}-d_{v,s_i})^2} \wedge 2m\\
\end{equation}
Since an arm $a_{v,s_i}$ is no longer pulled once either of the above conditions are reached, we have the total queries\footnote{If $v=w$, $T^p_{v,w}=0$} as
\begin{equation*}
    T^p_{v,s_i} \leq T^{A,p}_{v,s_i} \wedge T^{B,p}_{v,s_i}.
\end{equation*}

In both equations (\ref{eqn:blah1}) and (\ref{eqn:blah2}), we count the number of queries in each stage independent of the queries in the previous stage unlike in the algorithm implementation where information is preserved between different stages. We observe that an arm $a_{v,s_i}$ is pulled at most 
\begin{gather}
    T_{v,s_i} = \max\limits_{p=\{i, i+1, \dots, k-1\}} T^p_{v,s_i} \mbox{ } \forall v \in \mathcal{V}\backslash\mathcal{S} \nonumber\\
    T_{s_i,s_j} = \max\limits_{p=\{j, j+1, \dots, i-1\}} T^p_{s_i,s_j} \mbox{ } \forall v \in \mathcal{S}
\end{gather}
times. This is because if an arm is pulled these many times, it satisfies the confidence interval requirements at every stage $p$. Thus, summing over all pairs of $(v,s_i)$ and using (\ref{eqn:mdef}), we have the final result as stated in the statement of the theorem.

 \textit{Remark.} While the lower bound makes use of the physical structure of the data points, the upper bound doesn't take this into account. Hence, the upper bound can be made tighter by using the properties of the distance metric used (say the triangle inequality for a Euclidean space).

 \section{Solving (19)}\label{Sec:mgadetails}
In this section we present the details of computing $\vb*{\omega}^\star(\vb*{\mu})$. Without loss of generality, we consider $\mu_{i,1} \leq \mu_{i,2} \leq \dots \leq \mu_{i,b}$ for each $i$ and $\mu_{1,1} \geq \mu_{2,1} \geq \dots \geq \mu_{a,1}$, so that the maximin box $i^\star = 1$. The optimization problem (19) is a concave optimization problem over a convex constrained set. We solve this using the mirror gradient ascent algorithm with a negative entropy mirror map and simplex constraints imposed by taking a projection with respect to Bregman divergence \cite{Bubeck15}. As an example the numerical solution for an instance with $\vb*{\mu} \in \mathbb{R}^{3 \times3}$ with $\vb*{\mu} = [0.45\mbox{ }0.5\mbox{ }0.55;\mbox{ }0.35\mbox{ }0.4\mbox{ }0.6;\mbox{ }0.3\mbox{ }0.47\mbox{ }0.52]$ is $\vb*{\omega}^\star = [0.3633\mbox{ }, 0.1057\mbox{ }, 0.0532;\mbox{ }0.3738\mbox{ }\textbf{0}\mbox{ }\textbf{0};\mbox{ }0.1040\mbox{ }\textbf{0}\mbox{ }\textbf{0}]$.

Below, we present the details of the mirror gradient ascent algorithm used to solve the optimization problem (19). We state a supporting lemma and then proceed to optimizing the function to obtain our lower bound.
Define $\vb*{\lambda}$, belonging to the set of alternate instances as
\begin{align*}
    \mathcal{A}(\vb*{\mu}) & := \{\vb*{\lambda} \in \mathbb{R}^{a \times b} : i^\star(\vb*{\lambda}) \neq i^\star(\vb*{\mu}) \}. \nonumber
\end{align*}
Define function $f:\Tilde{\Omega}_{ab} \xrightarrow{} \mathbb{R}$ as 
\begin{align*}
f(\vb*{\omega}) & = \inf\limits_{\vb*{\lambda} \in \mathcal{A}(\vb*{\mu})} \sum\limits_i \sum\limits_j \omega_{i,j}kl_g(\mu_{i,j}, \lambda_{i,j}) \\ 
& = \min\limits_{i\neq 1, j} \mbox{ } \omega_{1,j}kl_g\left(\mu_{1,j}, \frac{\mu_{1,j}\omega_{1,j} + \mu_{i,1}\omega_{i,1}}{\omega_{1,j}+\omega_{i,1}}\right) \\
& {\color{white} ggggg} + \omega_{i,1}kl_g\left(\mu_{i,1}, \frac{\mu_{1,j}\omega_{1,j} + \mu_{i,1}\omega_{i,1}}{\omega_{1,j}+\omega_{i,1}}\right).
\end{align*}

Define
$$
\vb*{l}^\star(\vb*{\omega}) = kl_g(\vb*{\mu}, \vb*{\lambda}^{i^\star, j^\star}(\vb*{\omega}))
$$
where 
\begin{align*}
    (i^\star, j^\star)(\vb*{\omega}) =  \argmin\limits_{i \neq 1,j} \mbox{ }  & \omega_{1,j}kl_g(\mu_{1,j}, \frac{\mu_{1,j}\omega_{1,j} + \mu_{i,1}\omega_{i,1}}{\omega_{1,j}+\omega_{i,1}})\\
    & + \omega_{i,1}kl_g(\mu_{i,1}, \frac{\mu_{1,j}\omega_{1,j} + \mu_{i,1}\omega_{i,1}}{\omega_{1,j}+\omega_{i,1}}),
\end{align*}
and $\vb*{\lambda}^{i^\star, j^\star}$ same as $\vb*{\mu}$ except in two positions $\vb*{\lambda}^{i^\star, j^\star}(1,j^\star) = \vb*{\lambda}^{i^\star, j^\star}(i^\star, 1) = \frac{\mu_{1,j^\star}\omega_{1,j^\star} + \mu_{i^\star,1}\omega_{i^\star,1}}{\omega_{1,j^\star}+\omega_{i^\star,1}}$ i.e.,
\begin{align*}
\vb*{\lambda}^{i^\star, j^\star} = [& \mu_{1,1}, \dots, \mu_{1, j^\star -1}, \frac{\mu_{1,j^\star}\omega_{1,j^\star} + \mu_{i^\star,1}\omega_{i^\star,1}}{\omega_{1,j^\star}+\omega_{i^\star,1}},\\
& \mu_{1, j^\star + 1}, \dots, \mu_{1,b}; \dots, \mu_{i-1, b};\\
& \frac{\mu_{1,j^\star}\omega_{1,j^\star} + \mu_{i^\star,1}\omega_{i^\star,1}}{\omega_{1,j^\star}+\omega_{i^\star,1}}, \mu_{i, 2}, \dots, \mu_{i,b};\\
& \dots, \mu_{a,b}].
\end{align*}

\textbf{Lemma 4:} (\cite{GarivierGT}, Lemmas 7 \& 8) Let $\mathcal{A}(\vb*{\mu}) \subseteq \mathbb{R}^{a\times b}$ be a compact set. The function $f$ is a concave function and $\vb*{l}^\star(\vb*{\omega})$ is a supergradient of $f$ at $\vb*{\omega}$. $f$ is also $L$-Lipschitz with respect to $\lVert \cdot \rVert_1$ for any $L \geq \max\limits_{i,j} kl_g(\mu_{v_i,s_i}, \mu_{v_j,s_j})$.

Our optimization problem (19) is hence, 
\begin{equation*}\label{eqn:gaobj}
    T^\star(\vb*{\mu})^{-1} = \max\limits_{\vb*{\omega} \in \Tilde{\Omega}_{a,b}} f(\vb*{\omega}).
\end{equation*}

Since we can compute the supergradient of the concave function $f$ from Lemma 4, any supergradient ascent based algorithm can be used to find the solution $\vb*{\omega}^\star = \argmax\limits_{\vb*{\omega} \in \Tilde{\Omega}_{ab}} f(\vb*{\omega})$ to our optimization problem. Here, we use the mirror gradient ascent algorithm with the negative entropy function as the mirror map $\vb*{\Phi}(\vb*{x}) = \sum\limits_i x_i \log x_i$. We apply simplex constraints by taking a projection with respect to Bregman divergence. If the learning rate is represented by $\eta$, the gradient ascent equation in iteration $z$ can be written as 
\begin{align*}\label{eqn:gradientascent}
    \vb*{\omega}'_{z+1} & \leftarrow \vb*{\omega}_{z} e^{\eta \vb*{l}^\star(\vb*{\omega}_{z})} \nonumber\\
    \vb*{\omega}_{z+1} & \leftarrow \frac{\vb*{\omega}'_{z+1}}{\sum\limits_i \sum\limits_j (\omega'_{z+1})_{i,j}}.
\end{align*}
Note that $\forall i \geq 2, j\geq 2$, $(\omega_z)_{v,s}$ remains $0$ in all iterations $z$ because both $(\omega_0)_{v,s} = 0$ and $(l^\star(\vb*{\omega}_{z}))_{v,s} = 0$. Hence, these elements aren't active in the gradient ascent algorithm but the vector shape and size is maintained to preserve the structure of the problem, aid explanation and easy implementation in code.

\textit{Theorem 5 (\cite{GarivierGT}, Theorem 9)} Let $(\omega_0) = 0$ $\forall i \geq 2, j\geq 2$ and $(\omega_0) = \frac{1}{a+b-1}$ else and learning rate $\eta = \frac{1}{L}\sqrt{\frac{2\log(a+b-1)}{z}}$. The function $f$ has the following guarantees 
$$
f(\vb*{\omega}^\star) - f\left(\frac{1}{z}\sum\limits_{i=0}^{z-1} \vb*{\omega}_i\right) \leq L\sqrt{\frac{2\log(a+b-1)}{z}}
$$
for any $L \geq \max\limits_{i,j} kl_g(\mu_{v_i,s_i}, \mu_{v_j,s_j})$.

\section{Proof of Theorem~4}\label{Sec:ProofThmTaSUB}
This proof is along the lines of Proposition 6 in \cite{garivier2016maximin} and Proposition 13 in \cite{kaufmann16a}. 

From Proposition 6 in \cite{garivier2016maximin}, we have
$$
\mathbb{P}_{\vb*{d}^p}(\tau_\delta^p < \infty, \argmax\limits_{v \in \mathcal{V}} \min\limits_{s \in \mathcal{S}_p} \hat{d}_{v,s}(\tau_\delta^p) \neq v^\star_p) \leq \delta.
$$
Thus, for every stage of finding a center, the algorithm finds the correct center with a probability $\geq 1-\delta$.

Now, to upper bound the number of samples, we first consider the event $\mathcal{E}$ $\forall v \in \mathcal{V}, \forall s \in \mathcal{S}_p$,
\begin{align*}
\mathcal{E} := \frac{t_{v,s}(t)}{t} & \xrightarrow{t \to \infty} \omega^\star_{v,s}(\vb*{d}^p) \\
\hat{\mu}_{v,s}(t) & \xrightarrow{t \to \infty} \mu_{v,s} .
\end{align*}
Due to the sampling strategy used and the law of large numbers, $\mathbb{P}_{\vb*{d}^p}(\mathcal{E})=1$. Now, on $\mathcal{E}$, $\exists t_0$ such that $\forall t \geq t_0$, $\hat{d}_{v^\star_p, s}(t) > \max_{v \neq v^\star_p} \hat{d}_{v, s^\star_v}(t)$ and 
\begin{align*}
    Z(t) & = \max\limits_{i \in \mathcal{V}} \min\limits_{i' \in \mathcal{V}\backslash\{i\}} \max\limits_{j' \in \mathcal{S}_p} \min\limits_{j \in \mathcal{S}_p} Z_{(i,j)(i',j')}(t) \\ 
    & = \min\limits_{v \neq v^\star_p} \min\limits_{s \in \mathcal{S}_p} t_{v^\star_p, s}kl_g(\hat{d}_{v^\star_p, s}, \rho_{s v}) + t_{v, s^\star_v}kl_g(\hat{d}_{v,s^\star_v}, \rho_{s v})\\
    & = t[\min\limits_{v \neq v^\star_p} \min\limits_{s \in \mathcal{S}_p} \frac{t_{v^\star_p,s}}{t}kl_g(\hat{d}_{v^\star_p, s}, \rho_{s v}) {\color{white}}\\
    & {\color{white}safgsdfgsdfgfgsdfgsdfsddf}+ \frac{t_{v, s^\star_v}}{t}kl_g(\hat{d}_{v,s^\star_v}, \rho_{s v})]
\end{align*}
where $\rho_{sv} = \frac{t_{v^\star_p,s}\hat{d}_{v^\star_p,s} + t_{v,s^\star_v}\hat{d}_{v,s^\star_v}}{t_{v^\star_p,s}+t_{v,s^\star_v}}$.
For $\epsilon > 0$, $\exists t_1 \geq t_0$ such that $\forall t \geq t_1$ and $\forall v\neq v^\star_p$, 
\begin{align*}
& \frac{t_{v^\star_p,s}}{t}kl_g(\hat{d}_{v^\star, s}, \rho_{s v}) + \frac{t_{v, s^\star_v}}{t}kl_g(\hat{d}_{v,s^\star_v}, \rho_{s v}) \geq \\
& \frac{1}{1+\epsilon}[\omega^\star_{v^\star_p,s}kl_g(\hat{d}_{v^\star, s}, \rho_{s v }) + \omega^\star_{v,s^\star_v}kl_g(\hat{d}_{v,s^\star_v}, \rho_{s v })].
\end{align*}
Hence, for $t \geq t_1$,
\begin{align*}
Z(t) & \geq \frac{t}{1+\epsilon}[\min\limits_{v \neq v^\star_p} \min\limits_{s \in \mathcal{S}_p} \omega^\star_{v^\star_p, s}kl_g(\hat{d}_{v^\star_p, s}, \rho_{s v}) \\
& {\color{white}safssdfgsdfgdfgsdfgsdfgf} + \omega^\star_{v,s^\star_v}kl_g(\hat{d}_{v,s^\star_v}, \rho_{s v})].
\end{align*}
Consequently, for some positive constant $c$,
\begin{align*}
    \tau_\delta^p & = \inf \{t \in \mathbb{N} : Z(t) \geq \beta(t,\delta)\}\\
    & \leq t_1 \vee \inf\{t\in \mathbb{N}: t(1+\epsilon)^{-1}T^\star(\vb*{d}^p)^{-1} \geq \log(ct^\gamma/\delta)\}.
\end{align*}
Using Lemma 18 in \cite{kaufmann16a},  it follows on $\mathcal{E}$ for $\gamma \in [1,e/2]$,
\begin{align*}
    \tau_\delta^p & \leq t_1 \vee \gamma(1+\epsilon)T^\star(\vb*{d}^p)[\log(ce((1+\epsilon)T^\star(\vb*{d}^p))^\gamma/\delta) +\\
    & \mbox{\color{white} sdfgsdfgsdfgsdfgdshjhkjh} \log \log(c((1+\epsilon)T^\star(\vb*{d}^p))^\gamma/\delta) ].
\end{align*}
Thus $\tau_\delta^p$ is finite on $\mathcal{E}$ for every $\delta \in (0,1)$ and it can be upper bounded as
$$
\limsup\limits_{\delta \to 0} \frac{\tau_\delta^p}{\log(1/\delta)} \leq \gamma(1+\epsilon)T^\star(\vb*{d}^p).
$$
Summing over all stages of finding $k$ centers and letting $\epsilon \to 0$, we get the result as stated in the theorem.

\end{appendices}

\bibliography{mainpaper}

\end{document}